\documentclass[preprint]{revtex4}
\usepackage{amsmath}
\usepackage{amssymb}
\usepackage{natbib}
\usepackage{graphicx}
\usepackage{subfigure}
\usepackage{epsfig}
\usepackage{latexsym}
\usepackage{psfrag}
\usepackage{color}

\newcommand\grad{\mbox{\boldmath{$\nabla$}}}
\newcommand\lapl{\mbox{$\Delta$}}

\newcommand\boldr{\mathbf{r}}
\newcommand\boldrp{\mathbf{r'}}
\newcommand\dd{\text{d}}
\newcommand\kB{k_{\text{B}}}
\newcommand\bb{{\bar\beta}}

\definecolor{Red}{cmyk}{0,1,1,0}
\definecolor{Green}{cmyk}{1,0,1,0}
\begin{document}

\title{Equilibrium gas-liquid-solid contact angle from density-functional theory}

\author{Antonio Pereira}
\author{Serafim Kalliadasis}
\affiliation{Department of Chemical Engineering, Imperial College London\\
London, SW7 2AZ, United Kingdom}

\begin{abstract}
We investigate the equilibrium of a fluid in contact with a solid
boundary through a density-functional theory. Depending on the
conditions, the fluid can be in one phase, gas or liquid, or two
phases, while the wall induces an external field acting on the fluid
particles. We first examine the case of a liquid film in contact
with the wall. We construct bifurcation diagrams for the film
thickness as a function of the chemical potential. At a specific
value of the chemical potential, two equally stable films, a thin
one and a thick one, can coexist. As saturation is approached, the
thickness of the thick film tends to infinity. This allows the
construction of a liquid-gas interface that forms a well defined
contact angle with the wall.

\vspace{3mm}

\noindent {\it Keywords}: contact angle; density-functional theory.
\end{abstract}

\maketitle
\section{Introduction}

Wetting phenomena have received considerable attention over the past
few decades (comprehensive and detailed reviews are given
by \cite{RevModPhys_81_2009} and \cite{RevModPhys_57_1985}). Central
to any description of wetting is the presence of a contact line
which involves a three-phase conjunction (gas-liquid-solid). In the
case of dynamic wetting the associated moving contact line problem
is characterized by a stress singularity at the contact line. As a
consequence, formulating this problem in the framework of
conventional fluid mechanics leads to fundamental difficulties
concerning the modeling of the contact line
region \cite[]{JFluidMech_18_1963,JCollInt_35_1971,JFluidMech_65_1974}.
Accordingly, a variety of models have been proposed to alleviate
these difficulties and to address the behavior of the associated
dynamic contact angle, with the most popular approaches being the
replacement of the no-slip condition with a slip model or the
elimination of the contact line all together with the use of a thin
precursor layer. More recently, new approaches/theories have
appeared, based e.g. on local diffuse-interface/Cahn-Hilliard-type
approaches which, in general, cannot be rigourously justified. Other
recent approaches include hybrid molecular-hydrodynamic simulations.
However, the question of how to rigorously coarse grain from the
micro- to the macro-scale and to accurately transfer molecular
information to the macro-scale has not been adequately addressed as
of yet.

The difficulties with the contact line region stem from the
multi-scale nature of the problem. At the macroscale the usual laws
of hydrodynamics apply. At the nanoscale intermolecular
interactions, often described by molecular dynamics/Monte Carlo
simulations, dominate. Yet, phenomena occurring at the nanoscale
often manifest themselves at the macroscale (where individual
molecules are not considered or equivalently the hydrodynamic laws
consider a very large number of molecules at the same time). Despite
drastic improvements in computational power, molecular simulations
are still only applicable for small fluid volumes.

A compromise between conventional hydrodynamics and molecular
simulations can be achieved by density-functional theory (DFT). On
the one hand, DFT is able to retain the microscopic details of a
macroscopic system at a computational cost significantly lower than
that used in molecular simulations. On the other hand, DFT is
rigorous compared to conventional phenomenological approaches. It is
applicable for both uniform and non-uniform (density exhibits
spatial variation) as well as confined systems (e.g. in the presence
of a wall) within a self-consistent theoretical framework provided
by the statistical mechanics of fluids. At the same time,
substantial progress has been made in recent years in the
development of realistic free-energy functionals which take into
account the thermodynamic non-ideality attributed to the various
intermolecular forces. The papers
by \cite{AdvPhys_28_143}, \cite{PhysRevA_31_1985} and
\cite{AmJPhys_68_2000} outline a DFT framework currently widely
accepted by the statistical mechanics of fluids community and which
has also been used successfully in describing equilibrium
configurations in many different contexts, from fluids in pores to
liquid crystals, polymers and molecular self-assembly.

Several studies have examined the equilibrium of liquids in contact
with solids in different settings and in the framework of
DFT/statistical mechanics of fluids. For
example, \cite{ColloidsSurfacesA_206_11} adopted a DFT approach
based on a free-energy functional proposed by \cite{Landau}. This
study focused primarily on a one-dimensional (1D) configuration,
namely a liquid film in contact with a planar solid substrate but it
did discuss contact lines; for a example it gave an approximate
expression for the equilibrium contact angle obtained from the
sharp-interface limit of the DFT approach adopted and for large
distances from the solid substrate.
Recently, \cite{JChemPhys_129_2008_2,JChemPhys_129_2008_1,JChemPhys_130_2009}
adopted a DFT approach based on the framework described
by \cite{AdvPhys_28_143}, \cite{PhysRevA_31_1985}
and \cite{AmJPhys_68_2000} to calculate nanodrops on
chemically/physically inhomogeneous inclined/planar substrates and
to obtain the dependence of the contact angle of nanodrops on planar
horizontal substrates on the parameters of the intermolecular
interactions.

Our aim here is to provide a rigorous methodology for the treatment
of contact lines. Our approach is also based on the DFT framework
described by \cite{AdvPhys_28_143}, \cite{PhysRevA_31_1985}
and \cite{AmJPhys_68_2000}. We focus on the equilibrium of a fluid
in contact with a planar solid substrate whose understanding is
essential for the substantially more involved dynamics. We first
examine in detail the 1D case of a liquid film in contact with the
substrate. Particular emphasis is given on the bifurcation diagrams
for the film thickness as a function of the chemical potential. This
is a necessary step prior to understanding the more involved
two-dimensional (2D) case, i.e. the three-phase contact line. We
subsequently focus on the prewetting transition occurring at a
specific value of the chemical potential where two equally stable
films, a thin one and a thick one, coexist. The two thicknesses are
connected through a ridge-like interface which is sufficiently
smooth in the vicinity of the thin film for a long-range wall
potential but it becomes steep there when the wall potential is a
short-range one. When the co-existence value of the chemical
potential equals to the saturation one, the thickness of the thick
film tends to infinity. This is the case of a liquid wedge in
contact with the solid substrate and with a well-defined three-phase
contact line a few molecular diameters from the substrate. The wedge
seems to persist for all distances from the substrate. Hence, even
though DFT is a microscopic approach, it allows for the construction
of a macroscopic quantity such as contact angle. Moreover, unlike
macroscopic approaches e.g. Young's equation which naturally require
information on macroscopic parameters, such as the surface tensions
between the different phases, DFT relies only on first principles,
namely information related to intermolecular parameters. It also
elucidates the structure of the three-phase contact line region and
the precise role of fluid-fluid and fluid-solid (long-/short-range)
interactions there.

\section{Problem definition}
\label{pbdef}

\subsection{Setting}

\begin{figure}
\begin{center}
\includegraphics[width=0.8\linewidth]{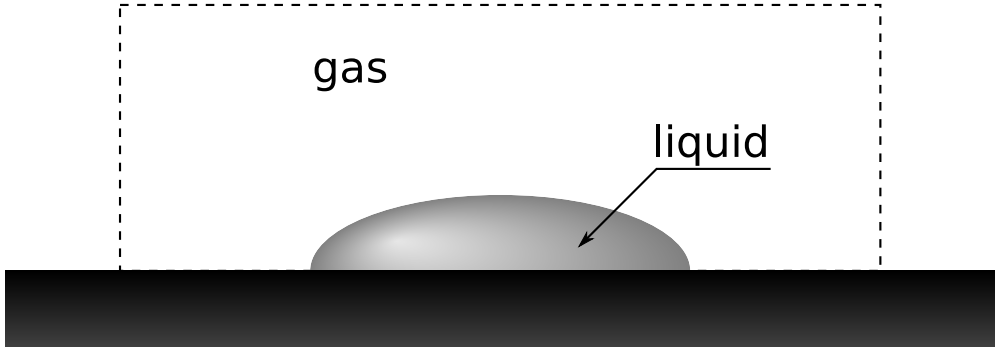}
\caption{\label{sketch}%
Sketch of the physical setting.}
\end{center}
\end{figure}

We consider part of a simple fluid inside a volume $\mathcal{V}$ in
contact with a horizontal planar substrate as shown in
figure~\ref{sketch}. The system, whose boundary is denoted with the
closed dotted line in the figure, is open, i.e. fluid particles can
come in and out, and its surroundings are at temperature $T$ and
chemical potential $\mu$. Depending on the conditions, the fluid can
be in one phase (gas or liquid) or two phases. The fluid particles
interact through a pair potential $\phi(r_{12})$, where $r_{12}$ is
the distance between the centers of mass of the particles. The main
effect of the substrate is to induce an external field $V(\boldr)$
acting on the fluid particles with $\boldr$ the position vector of
the inertial center of fluid particles. For the sake of simplicity,
gravity is neglected. A Cartesian coordinate system $(x,y,z)$ is
chosen such that $x$ and $y$ are parallel to the wall surface while
$z$ is the outward-pointing coordinate normal to the wall. Of
particular interest is the region around the solid/fluid interface.

\subsection{Density-functional theory of fluids}

The equilibrium state of the system is characterized by the fluid
density $n_0(\boldr)$, in units of number of particles per unit
volume (number density), defined as \cite[]{AdvPhys_28_143}:
\begin{equation}
n_0(\boldr)=<\sum_{i=1}^N\delta(\boldr-\boldr_i)>
\end{equation}
with $\boldr\in\mathcal{V}$. To obtain $n_0(\boldr)$, we use
elements from equilibrium statistical mechanics of fluids, in
particular DFT. It has been shown that there exists a functional
$\Omega[n]$, defined on a set of functions $n(\boldr)$ compatible
with the external potential $V(\boldr)$, which has the property to
be at a global minimum (for a given temperature, chemical potential
and external potential) when $n(\boldr)$ is equal to the equilibrium
density profile $n_0(\boldr)$ \cite[]{AdvPhys_28_143,Plischke}
(minimum principle). A second functional $F[n]$ can also be
introduced such that $\Omega[n]$ reads:
\begin{equation}
\Omega[n]=F[n]+\int\!\dd\boldr\,n(\boldr)(V(\boldr)-\mu) \text{,}
\label{omega}
\end{equation}
where the integral is understood to be a volume one over
$\mathcal{V}$. The intrinsic free energy and the grand potential of
the system at equilibrium are then equal to $F[n_0]$ and
$\Omega[n_0]$, respectively (such quantities can only be defined at
equilibrium).

By using variational calculus, it can be easily shown that a
necessary condition for a minimum of $\Omega[n]$ is:
\begin{equation}
\frac{\delta F[n]}{\delta n(\boldr)}+V(\boldr)=\mu.
\label{minimization}
\end{equation}
This equation is supplemented by appropriate boundary conditions,
i.e.\ the value of $n$ outside $\mathcal{V}$. In our case, the
external potential $V(\boldr)$ is due to the interaction between the
fluid and the wall. Note that since our system is open
(grand-canonical ensemble formalism), the total number of particles,
$N$, is not given, but it is obtained instead once $n_0(\boldr)$ is
known: $N = \int d\boldr n_0(\boldr)$. For simplicity, we drop the
subscript 0 from $n$ hereinafter.

\subsection{Fluid modeling}

Several approximations for $F[n]$ have been proposed over the years.
One which has proved to be successful in a number of cases is based
on a perturbation approach. The basic idea is to split the pair
potential $\phi(r_{12})$ into two terms:
$\phi(r_{12})=\phi_{\text{r}}(r_{12})+\phi_{\text{p}}(r_{12})$. The
first term, $\phi_{\text{r}}(r_{12})$, is a reference potential and
usually corresponds to the harshly repulsive part of $\phi(r_{12})$,
while the second term, $\phi_{\text{p}}(r_{12})$, acts as a
``perturbation" to $\phi_{\text{r}}(r_{12})$ and is in general a
slowly attractive potential. Such separation, which is reminiscent
of the approach behind the van der Waals equation of state for gases
and can thus be viewed as a generalization of this approach, is in
general well-suited for treating dense fluids. Its underlying
physical motivation is that the structure of a fluid is determined
mainly by the repulsive forces. After some additional
simplifications we have the following compact expression for $F[n]$
\cite[]{AdvPhys_28_143,Plischke}:
\begin{subequations}
\begin{equation}
F[n]=F_{\text{r}}[n]+\frac{1}{2}%
\int\!\!\!\!\int\!\dd\boldr\dd\boldrp\;n(\boldr)n(\boldrp)%
\phi_{\mathrm{p}}(|\boldr-\boldrp|)
\text{.}
\label{fngeneral}
\end{equation}
The first term of equation~(\ref{fngeneral}) accounts for
$\phi_{\text{r}}(r_{12})$ and can be further simplified by using the
co-called local density approximation:
\begin{equation}
F_{\text{r}}[n]=\int\!\dd\boldr n(\boldr)f(n(\boldr))
\text{,}
\label{local}
\end{equation}
\end{subequations}
where $f$ is the local free energy per particle of the reference
fluid. The main simplification used to obtain the second term of
equation~(\ref{fngeneral}), which involves the attractive part of the
interaction potential $\phi(r_{12})$, is to express the pairwise
distribution function, i.e. the probability that two particles
occupy the positions $\boldr_1$, $\boldr_2$, as the product of
$\eta(\boldr_1)$, $\eta(\boldr_2)$ and the radial distribution
function of the uniform reference fluid.

The decomposition of the pair potential in two parts can be done in
different ways. In the Barker-Henderson approach for
example \cite[]{JChemPhys_47_4714}, the attractive part is,
\begin{equation}
\phi_{\text{p}}(r_{12})={}
\begin{cases}
\displaystyle
0\quad&\text{if $r_{12}<\sigma$} \\
\displaystyle
4\epsilon\left[{\left(\frac{\sigma}{r_{12}}\right)}^{12}%
-{\left(\frac{\sigma}{r_{12}}\right)}^6\right]%
\quad&\text{if $r_{12}>\sigma$,}
\end{cases}
\end{equation}
where $\epsilon$ is a parameter that measures the strength of the
potential. Another possibility is the decomposition suggested
by \cite{JChemPhys_54_5237}. Here we adopt the Barker-Henderson
approach. The reference potential is then,
$\phi_{\text{r}}=\phi-\phi_{\text{p}}$. However, to take full
advantage of the perturbation technique, the reference system should
be a well-known fluid: the preferred choice is often a hard-sphere
fluid, whose bulk local free energy is well described by the
Carnahan-Starling expression:
\begin{subequations}
\begin{equation}
\beta f(T,n)=\ln(\Lambda^3 n)-1%
+\frac{\eta(4-3\eta)}{{(1-\eta)}^2}
\label{carnahanstarling}
\end{equation}
where
\begin{equation}
\eta\equiv\frac{\pi}{6}d^3n
\text{,}
\label{packing}
\end{equation}
\end{subequations}
is the packing fraction, $d$ is the hard-sphere diameter, $\Lambda$
is the thermal de Broglie length and $\beta\equiv 1/(\kB T)$ with
$\kB$ the Boltzmann constant. This means that $\phi_{\text{r}}$ is
in fact approximated by a hard-sphere potential. The associated
molecular diameter $d$, which appears in equation~(\ref{packing})
and hence~(\ref{carnahanstarling}), can be linked to $\sigma$ and
the reference part of the interaction
potential~\cite[]{JChemPhys_47_4714} but for simplicity we assume
$d=\sigma$.

Instead of equations~(\ref{local}) and (\ref{carnahanstarling}), one
could use a more refined approach for a hard sphere fluid based on
Rosenfeld's fundamental measure theory \cite[]{PhysRevLett_63_980}.
In general, this approach gives better results for the fluid density
at small distances from the wall (a few molecular diameters) but
requires a more involved computational treatment. Our aim here is to
keep the formalism as simple as possible and yet retain the basic
ingredients of the underlying physics.

By substituting equations~(\ref{fngeneral}) and~(\ref{local}) into
equation~~(\ref{minimization}), we obtain:
\begin{subequations}
\label{integralequationthreedall}
\begin{equation}
\mu_{\text{r}}(n)+\int\!\dd\boldrp n(\boldrp)%
\phi_{\mathrm{p}}(|\boldr-\boldrp|)+V(\boldr)=\mu
\label{integralequationthreed}
\end{equation}
where
\begin{equation}
\mu_{\text{r}}(n)\equiv{\left(\frac{\partial(nf(T,n))}{\partial
n}\right)}_T\text{,}
\end{equation}
\end{subequations}
is the chemical potential of the reference system which in the case
of a hard sphere fluid and using equation~(\ref{carnahanstarling})
reads:
\begin{equation}
\beta\mu_{\text{r}}(n)=\ln\left(\Lambda^3n\right)+%
\frac{\eta(8-9\eta+3\eta^2)}{{(1-\eta)}^3}
\text{.}
\end{equation}

When the external potential vanishes and the fluid is uniform, i.e.
its density does not exhibit any spatial variation,
equation~(\ref{fngeneral}) combined with equation~(\ref{local}),
reduces to
\begin{subequations}
\begin{equation}
F[n]=\mathcal{V}n\bar{f}(n)
\end{equation}
where
\begin{equation}
\bar{f}(n)=f(n)-\alpha n \quad \mbox{and} \quad
\alpha=-\frac{1}{2}\int_{-\infty}^{+\infty}\!\!\!\dd\mathbf{r}\,\phi_{\text{p}}(|\mathbf{r}|)
\text{.}
\end{equation}
\end{subequations}
Equation~(\ref{integralequationthreedall}) is then an algebraic
equation with parameters $T$ and $\mu$. Using for $f$ the expression
given in equation~(\ref{carnahanstarling}), the usual thermodynamics
for liquid/gas systems applies. The solutions to
equation~(\ref{integralequationthreedall}) correspond to extrema of
the functional $\Omega[n]$. There is one solution for
$T>T_{\text{c}}$ and one to three solutions for $T<T_{\text{c}}$,
where $T_{\text{c}}$ is the critical temperature
($\beta_{\text{c}}\equiv 1/(\kB T_{\text{c}}))$. In the latter case,
there are two minima of equal depth for $\Omega$, at small and large
densities, respectively, and a maximum at intermediate densities.
The middle solution is unstable while the other two correspond to
the liquid and gas bulk densities, denoted as
$n_{\text{gas}}(T,\mu)$ and $n_{\text{liq}}(T,\mu)$, respectively,
with $n_{\text{gas}}(T,\mu)<n_{\text{liq}}(T,\mu)$. The chemical
potential at the gas/liquid transition will be denoted as
$\mu_{\text{sat}}(T)$. The gas is the preferred state for $\mu <
\mu_{\text{sat}}(T)$. At $\mu = \mu_{\text{sat}}(T)$ the gas and
liquid bulk are at equilibrium while for $\mu > \mu_{\text{sat}}(T)$
the liquid is the preferred state. Note that in all cases considered
here $T < T_{\text{c}}$ as close to the critical temperature thermal
fluctuations, which are not taken into account in the DFT formalism
adopted here, become significant.

\subsection{Wall potentials}

Three types of wall potential are considered in this study. All are
attractive at large distances. The first two have the same infinite
repulsive part but they differ in the range of the attraction term.
The short-range potential reads,
\begin{subequations}
\begin{equation}
V_{\text{SR}}(z)={}%
\begin{cases}
\displaystyle
\;+\infty\quad&\text{if $z<z_{\text{w}}$}\\
\displaystyle
\;-\epsilon_{\text{w}}\exp\left(-\frac{z-z_{\text{w}}}{\sigma_{\text{w}}}\right)\quad&\text{if
$z>z_{\text{w}}$}
\end{cases}
\label{vextexp}
\end{equation}
while for the long-range one we use,
\begin{equation}
V_{\text{LR}}(z)={}%
\begin{cases}
\;+\infty\quad&\text{if $z<z_{\text{w}}$}\\
\displaystyle
-2\epsilon_{\text{w}}{\left(\frac{1}{1+\frac{z-z_{\text{w}}}{\sigma_{\text{w}}}}\right)}^3\quad&\text{if
$z>z_{\text{w}}$}.
\end{cases}
\label{vextalg}
\end{equation}
In these expressions, $\epsilon_{\text{w}}$, $\sigma_{\text{w}}$ and
$z_{\text{w}}$ are three wall parameters; the first two are strictly
positive and are related to the strength and range of the
potentials. The expressions in equations \ref{vextexp} and
\ref{vextalg} are useful in that they allow us to examine the effect
of the wall attraction, and in particular its range, on the fluid
equilibrium state. The last wall potential has a smooth repulsive
part compared to the previous ones (i.e. it is non-infinite):
\begin{equation}
V_{\text{LJ}}(z)\!=\!\!{}%
\begin{cases}
\;+\infty&\text{if $z<z_{\text{w}}$}\\
\displaystyle
2\epsilon_{\text{w}}\!\!\left[\frac{2}{15}{\left(\frac{\sigma_{\text{w}}}{z-z_{\text{w}}}\right)\!}^9%
\!\!-{\left(\frac{\sigma_{\text{w}}}{z-z_{\text{w}}}\right)\!}^3\!\right]\!\!\!\!&\text{if
$z>z_{\text{w}}$}.
\end{cases} \label{vextlj}
\end{equation}
\end{subequations}
It can be derived by considering that the wall is made of a uniform
density of particles interacting with the fluid particles through a
Lennard-Jones potential (see Appendix A). By setting
$\sigma_{\text{w,LR}}=z_{\text{w,LR}}-z_{\text{w,LJ}}$ and
$\epsilon_{\text{w,LR}}=(\sigma_{\text{w,LJ}}/\sigma_{\text{w,LR}})^3%
\epsilon_{\text{w,LJ}}$, the two last potentials have the same
attractive part. Finally, it is also possible to have a purely
repulsive wall with no attraction (a hard or dry wall),
\begin{equation}
V(z)={}%
\begin{cases}
\;+\infty\quad&\text{if $z<z_{\text{w}}$}\\
\;0\quad&\text{if $z>z_{\text{w}}$}. \nonumber
\end{cases}
\end{equation}

\section{1D problem}
\label{onedpb}

\subsection{Equations}

In this section we assume that the system is invariant along the $x$
and $y$ directions and infinite in the $z$-direction. Equation
(\ref{integralequationthreed}) then becomes:
\begin{subequations}
\label{1dintegralequation}
\begin{equation}
\mu_{\text{r}}(n)+\int_{-\infty}^{+\infty}\!\dd z' n(z')%
\Phi_{\text{1d}}(z'-z)+V(z)=\mu
\label{integralequationoned}
\end{equation}
where
\begin{equation}
\label{1dsimpleintegral}
\Phi_{\text{1d}}(Z)=\int\!\!\!\!\!\int_{-\infty}^{+\infty}\!\!\!\dd x\,\dd y\,%
\phi_{\text{p}}\!\left(\sqrt{x^2+y^2+Z^2}\right); \quad Z \equiv z' - z.
\end{equation}
\end{subequations}

The integral in~(\ref{1dsimpleintegral}) can be performed
analytically (see Appendix A). To solve numerically the integral
equation~(\ref{1dintegralequation}) a simple iterative procedure can
be employed based on a Picard scheme. The integral is first computed
with the values of the density obtained from the previous iteration
by using a simple trapezoidal rule and then the remaining terms are
solved for $n$, which gives the new value for the density. We found
that, with the exception of unstable solutions, this scheme is
robust and allows the use of an initial guess which can be far from
the solution. However, convergence tends to be slow after the first
iterations. This can be greatly improved by using instead a more
involved scheme based on a Newton procedure in which the Jacobian
matrix is appropriately simplified. Details are given in Appendix B.

\subsection{Density profiles}

\begin{figure}
\begin{center}
\includegraphics[width=0.8\linewidth]{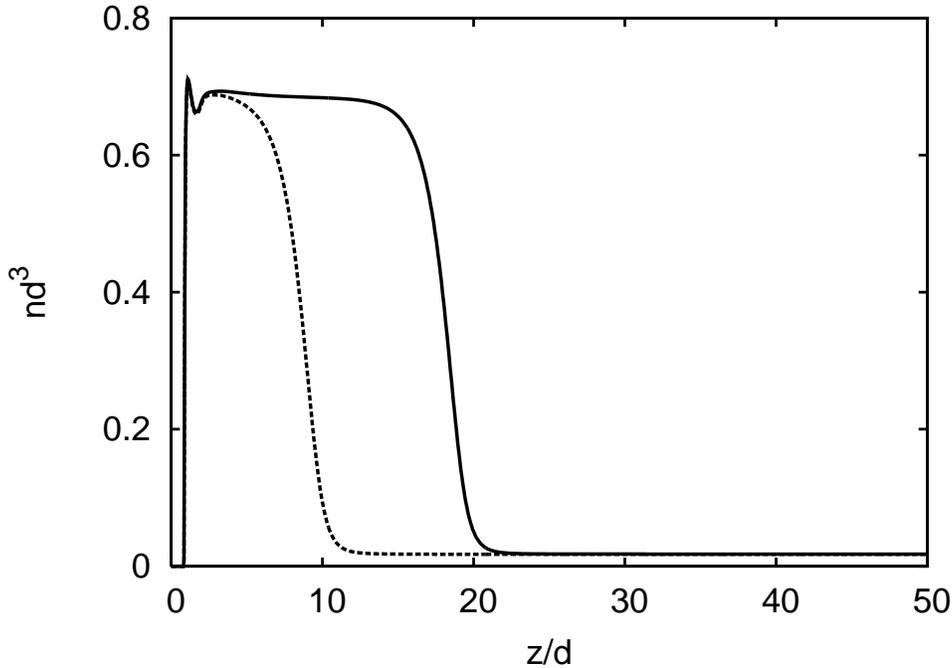}
\caption{\label{wallgas}%
Density profiles for $\mu<\mu_{\text{sat}}$ at $T=0.7\,T_{\text{c}}$
near an attractive wall. The wall potential is $V_{\text{LJ}}$ given
by equation~(\ref{vextlj}) with parameters
$\beta_{\text{c}}\epsilon_{\text{w}}=1.8$,
$\sigma_{\text{w}}=1.25\,d$ and $z_\text{w}=0$. Solid line,
$\mu-\mu_{\text{sat}} = -0.001 \kB T_{\text{c}}$; dotted line,
$\mu-\mu_{\text{sat}} = -0.01 \kB T_{\text{c}}$. In the former case,
$n_{\text{liq}}d^3\approx0.684$ which is close to the value of the
density for $z/d$ in the interval $[3,17]$.}
\end{center}
\end{figure}

In figures~\ref{liqgas} and~\ref{wallgas}, we present typical
profiles obtained by solving equation~(\ref{1dintegralequation}).
Figure \ref{liqgas} depicts the density profile of the fluid along
the $z$-direction when $\mu=\mu_{\text{sat}}(T)$ in the absence of
the wall. For this value of $\mu$, both bulk liquid and bulk gas are
equally possible states ($\Omega$ has two variational minima of
equal ``depth" at $n_{\text{liq,gas}}$). The interface region is
just a few molecular diameters thick. The location of the sharp
interface, say $z_\text{I}$, can be obtained from the Gibbs dividing
surface defined from $\int_{-\infty}^{z_{\text{I}}}
(n_{\text{liq}}-n)\dd z = \int_{z_{\text{I}}}^{+\infty}
(n-n_{\text{gas}})\dd z$.

Figure~\ref{wallgas} shows profiles obtained when the wall is
switched on and exerts a long-range attraction on the fluid
molecules. In this case, $\mu<\mu_{\text{sat}}(T)$ so that the bulk
gas is the more stable phase in the absence of the wall ($\Omega$
has a single variational minimum at $n_{\text{gas}}$). As
$z\rightarrow+\infty$ the density tends to the density of the gas,
$n_{\text{gas}}$, but near the wall the attractive potential induces
a bump in the density profile. If $\mu$ is close enough to
$\mu_{\text{sat}}(T)$ (solid line in figure~\ref{wallgas}), this
bump is more pronounced while the fluid density in that area is
roughly equal to the metastable liquid bulk density (corresponding
to a local variational minimum of $\Omega$). A thin liquid film is
effectively formed between the wall and the gas. It can only exist
due to the attraction with the wall, in other words an attractive
wall stabilizes a thin liquid film in contact with it. We also note
the small oscillations in the density profile near the wall
corresponding to adsorption of the liquid particles there. More
pronounced oscillations would have been found in this area if the
reference system of hard sphere fluid had been treated with a more
refined model like the Rosenfeld fundamental measure theory
mentioned earlier \cite[]{PhysRevLett_63_980}. Also, when a thin
film is present, we observe that the interface between the film and
the gas phase is very similar (although $\mu$ is different) to the
profile given in figure~\ref{liqgas} suggesting that in that case,
the wall has little influence. In fact, as $\mu \rightarrow
\mu^{-}_{\text{sat}}$, the thickness of the liquid film tends to
infinity and we approach the case depicted in figure~\ref{liqgas},
i.e. it is like the wall is not even present (with the exception of
course of the area close to it where the density oscillations
occur).

\begin{figure}
\begin{center}
\includegraphics[width=0.8\linewidth]{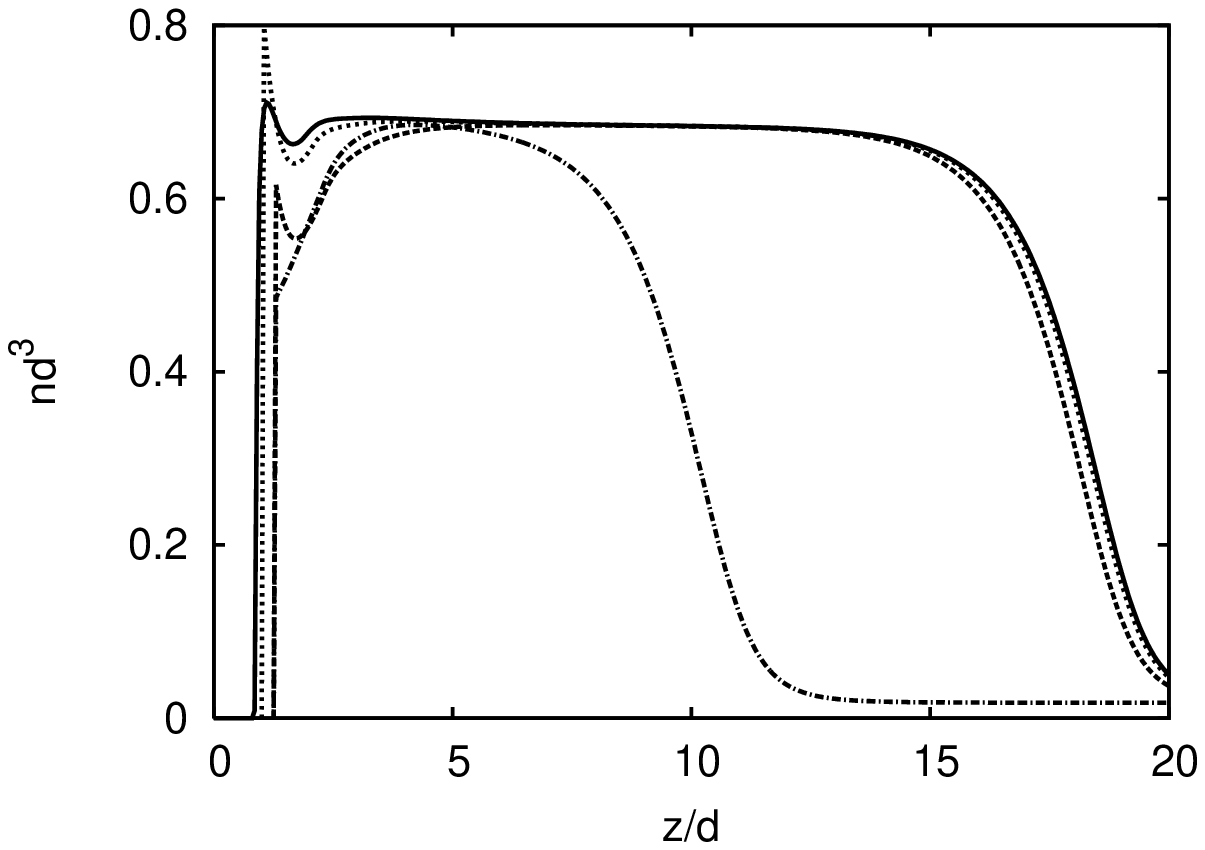}
\caption{\label{wallgasparam}%
Density profiles for $\mu-\mu_{\text{sat}}-0.001 \kB T_{\text{c}}$
at $T=0.7\,T_{\text{c}}$ near an attractive wall.
Solid line:
wall potential $V_{\text{LJ}}$ with $\beta_{\text{c}}\epsilon_{\text{w}}=1.8$,
$\sigma_{\text{w}}=1.25\,d$ and $z_\text{w}=0$;
dashed line:
wall potential $V_{\text{LR}}$ with $\beta_{\text{c}}\epsilon_{\text{w}}=1.8$,
$\sigma_{\text{w}}=1.25\,d$ and $z_\text{w}=\sigma_{\text{w}}$;
dotted line:
wall potential $V_{\text{LR}}$ with $\beta_{\text{c}}\epsilon_{\text{w}}\approx3.52$,
$\sigma_{\text{w}}=d$ and $z_\text{w}=\sigma_{\text{w}}$;
dashed-dotted line:
wall potential $V_{\text{SR}}$ with $\beta_{\text{c}}\epsilon_{\text{w}}=1.8$,
$\sigma_{\text{w}}=1.25\,d$ and $z_\text{w}=\sigma_{\text{w}}$.}
\end{center}
\end{figure}

For given external conditions of temperature and chemical potential,
an illustration of the dependance of the density profile on the
characteristics of the wall potential is shown in figure
\ref{wallgasparam}. We first note when comparing wall potentials
$V_{\text{SR}}$ (dashed-dotted line) and $V_{\text{LR}}$ (dashed
line), that the liquid film thickness in the first case is
considerably smaller even though all wall parameters are identical
and in particular the integrals between $z_{\text{w}}$ and $+\infty$
of the two wall potentials are the same. In contrast, when
$z_{\text{w,LR}}=z_{\text{w,LJ}}+\sigma_{\text{w,LR}}$, wall
potentials $V_{\text{LR}}$ (dashed line) and $V_{\text{LJ}}$ (solid
line) yield similar profiles except near the wall. These two last
profiles can be made even closer to each other, including the area
close to the wall, by changing the value of $z_{\text{w,LR}}$
through $\sigma_{\text{w,LR}}$ (dotted line) and updating
$\epsilon_{\text{w,LR}}$ accordingly. However, even if the profiles
can become similar, important differences can still exist in the
adsorption isotherms (cf next section).

\begin{figure}
\begin{center}
\includegraphics[width=0.8\linewidth]{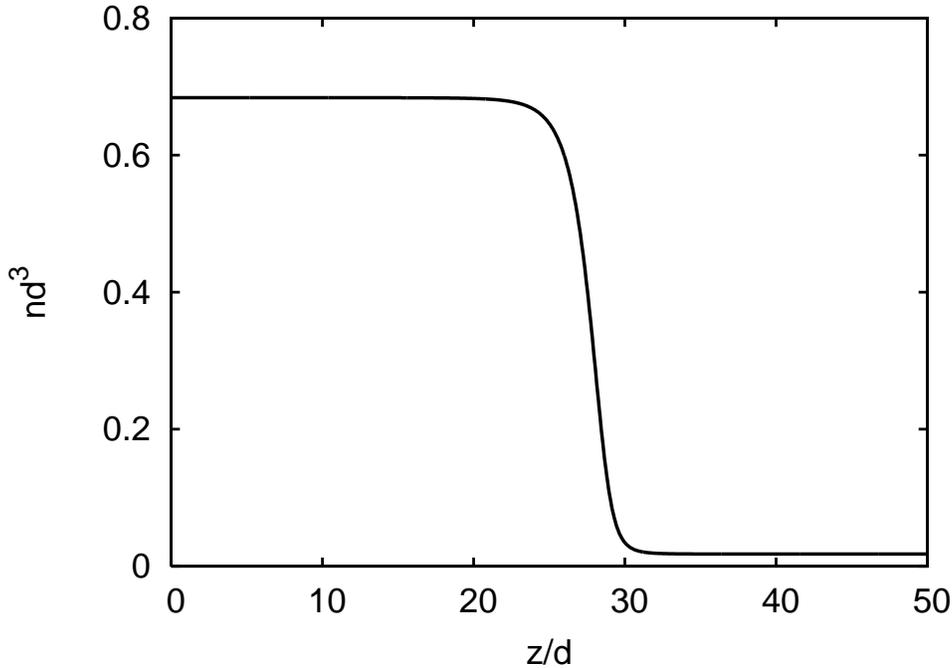}
\caption{\label{liqgas}%
Density in the vicinity of a liquid-gas interface for
$\mu=\mu_{\text{sat}}$ at $T=0.7\,T_{\text{c}}$ in the absence of
the wall. The domain size used in the computation is [0-50] and the
number of points is 1000. The bulk liquid density is
$n_{\text{liq}}d^3\approx0.684$ and the bulk gas one is
$n_{\text{gas}}d^3\approx0.0178$.}
\end{center}
\end{figure}

\subsection{Isotherms}
\label{1disotherms}

Of particular interest is the dependence of the film thickness with
respect to $\mu$ for a given $T$. This is crucial to understanding
more involved configurations such as the contact angle case examined
later. Because of the spatial variation of the density, it is in
fact more relevant to consider an integral norm of the density, the
``adsorption", defined as
\begin{equation}
\Gamma=\int_{-\infty}^{z_{\text{I}}}\dd z\,n(z)+\int_{z_{\text{I}}}^{+\infty}\dd z\,(n(z)-n_{\infty})
\end{equation}
where $z_{\text{I}}$ is the position of the dividing wall-fluid
interface and $n_{\infty}$ is in the present case equal to
$n_{\text{gas}}$ ($n=0$ as $z\rightarrow-\infty$).
The bifurcation diagrams for the isotherms are constructed with a
continuation procedure which also involves an extra
e\-qua\-tion/ge\-ome\-tric constraint which links the continuation
parameter $\mu$ and $n$ with the continuation step. Besides easily
obtaining the density profiles for a wide range of values of $\mu$,
this approach has the additional benefit that it allows the
computation of unstable branches, which would have been more
difficult if not impossible to obtain otherwise.

\begin{figure}
\begin{center}
\includegraphics[width=0.8\linewidth]{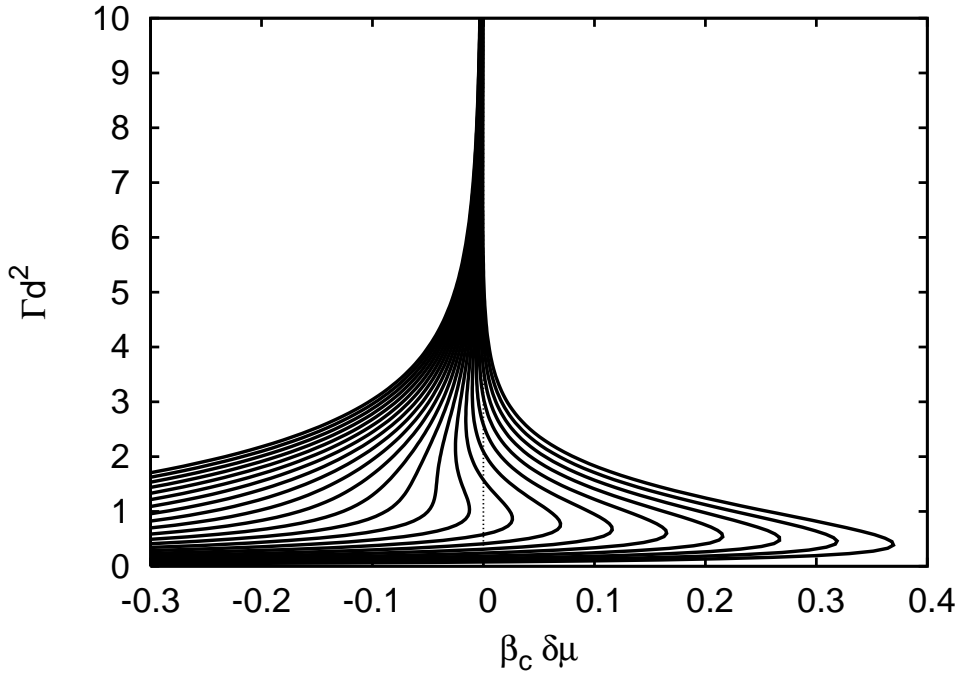}
\caption{\label{biflj}%
Bifurcation diagram for $\Gamma$ as a function of $\delta\!\mu
\equiv \mu-\mu_{\text{sat}}$, for the attractive wall with potential
$V_{\text{LJ}}$ (equation~(\ref{vextlj})), $z_{\text{w}}=0$ and
$\sigma_{\text{w}}=1.25\,d$. The temperature is
$T=0.7\,T_{\text{c}}$ and the dividing interface is located at
$z_{\text{I}}=z_{\text{w}}$. The wall parameter
$\beta_{\text{c}}\epsilon_{\text{w}}$ varies from 0.8 (right curve)
to 3.0 (left curve) in steps of 0.1.}
\end{center}
\end{figure}

Typical isotherms for the wall potential $V_{\text{LJ}}$ in
equation~(\ref{vextlj}) are depicted in figure~\ref{biflj}. For very
attractive walls (for example, the case
$\beta_{\text{c}}\epsilon_{\text{w}}=3$ in figure~\ref{biflj}) and
$\mu<\mu_{\text{sat}}$, the adsorption $\Gamma$ is a monotonically
increasing function of $\mu$; in particular, only one solution
exists for a given value of $\mu$. At the saturation value
$\mu_{\text{sat}}$, the bulk liquid phase becomes as stable as the
bulk gas one and the adsorption goes to infinity. For less
attractive walls, a multi-valued S-shaped loop in the isotherm
appears for values of $\mu$ say in $[\mu_{-},\mu_{+}]$ and with
three branches of solutions from which the middle one is always
unstable (corresponding to variational maxima for $\Omega$). This
loop is associated with the presence of a first-order phase
transition with respect to the adsorption: for a certain value of
$\mu \in [\mu_{-},\mu_{+}]$, three solutions exist, of which those
in the lower and upper branches are equally stable (variational
minima of $\Omega$ of equal ``depth"), and consequently the system
can adopt for that value of $\mu$ two different adsorptions or
equivalently film thicknesses. This transition is often referred to
as the ``prewetting transition" and a Maxwell construction in the
variables $(\mu,\Gamma)$ can be carried out to find the coexistence
value $\mu_{\text{coex}}(T)$ of the chemical potential at which the
two states have the same stability. For $\mu \in
[\mu_{-},\mu_{\text{coex}}]$, the lower branch is stable and the
upper branch one metastable ($\Omega$ has two variational minima but
the one in the lower branch is ``deeper") while for $\mu \in
[\mu_{\text{coex}},\mu_{+}]$ the upper branch is stable and the
lower one metastable.

The value $\mu_{\text{sat}}$ imposes an upper bound on
$\mu_{\text{coex}}$. We cannot have $\mu_{\text{coex}} \geq
\mu_{\text{sat}}$ as a Maxwell construction in this region is not
possible. As $\mu_{\text{coex}} \rightarrow \mu^{-}_{\text{sat}}$ we
can have a Maxwell construction with a thin film equally stable with
an almost infinite thick film, while for $\mu = \mu_{\text{sat}}$
the most stable state is that of a thin film on the lower branch as
is the case when the wall is only slightly attractive (e.g.
$\beta_{\text{c}}\epsilon_{\text{w}}=0.8$ in figure~\ref{biflj}).
This is the signature of a partial wetting situation. Complete
wetting on the other hand occurs whenever $\mu_{\text{coex}} <
\mu_{\text{sat}}$ and as $\mu \rightarrow \mu_{\text{sat}}$ on the
upper branch. For a given value of $\epsilon_{\text{w}}$, one can
usually go from a partial wetting case to a complete wetting one by
changing the temperature.

\begin{figure}
\begin{center}
\includegraphics[width=0.8\linewidth]{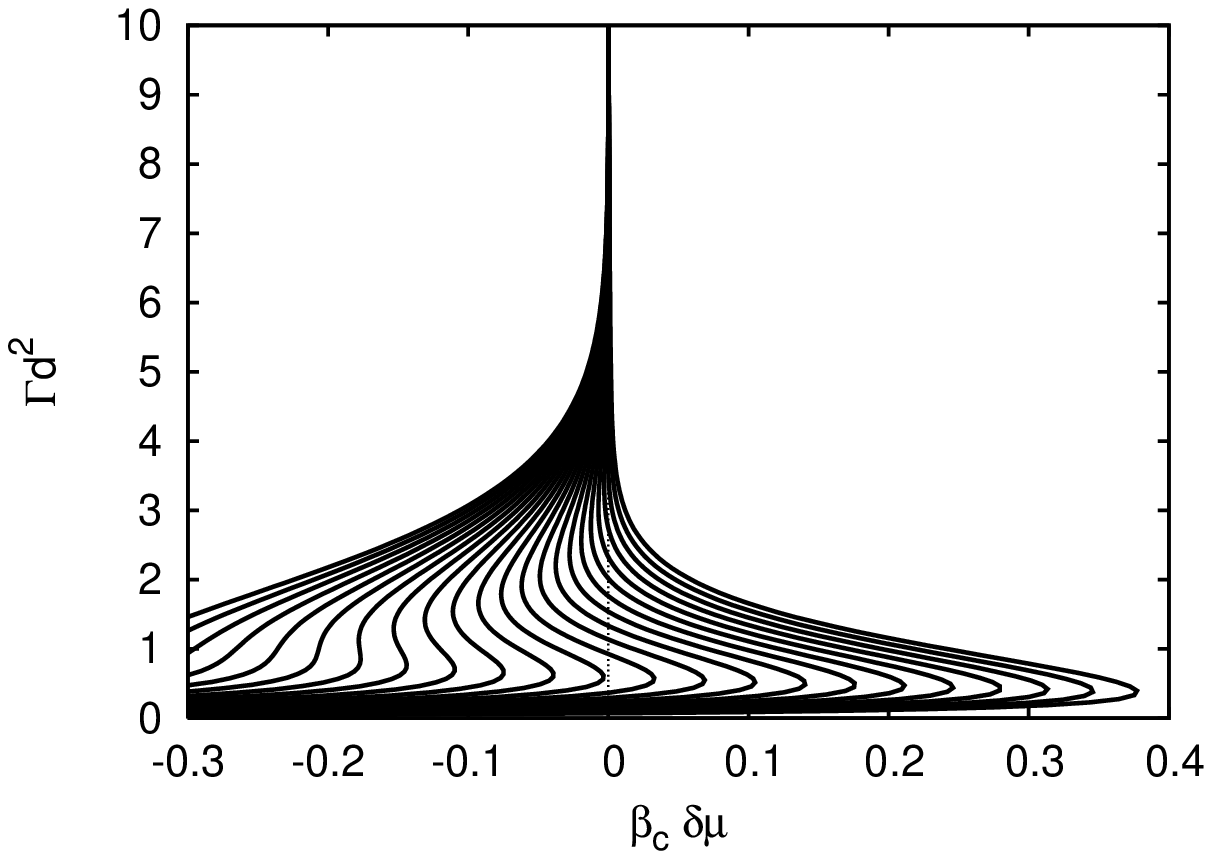}
\caption{\label{bifexp}%
Bifurcation diagram for $\Gamma$ as a function of
$\delta\!\mu\equiv\mu-\mu_{\text{sat}}$, for the attractive wall
with potential $V_{\text{SR}}$ (Eq.~(\ref{vextexp})),
$z_{\text{w}}=0$ and $\sigma_{\text{w}}=1.25\,d$. The temperature is
$T=0.7\,T_{\text{c}}$ and the dividing interface is located at
$z_{\text{I}}=z_{\text{w}}$. The wall parameter
$\beta_{\text{c}}\epsilon_{\text{w}}$ varies from 1.2 (right curve)
to 3.4 (left curve) in steps of 0.1.}
\end{center}
\end{figure}

\begin{figure}
\begin{center}
\includegraphics[width=0.8\linewidth]{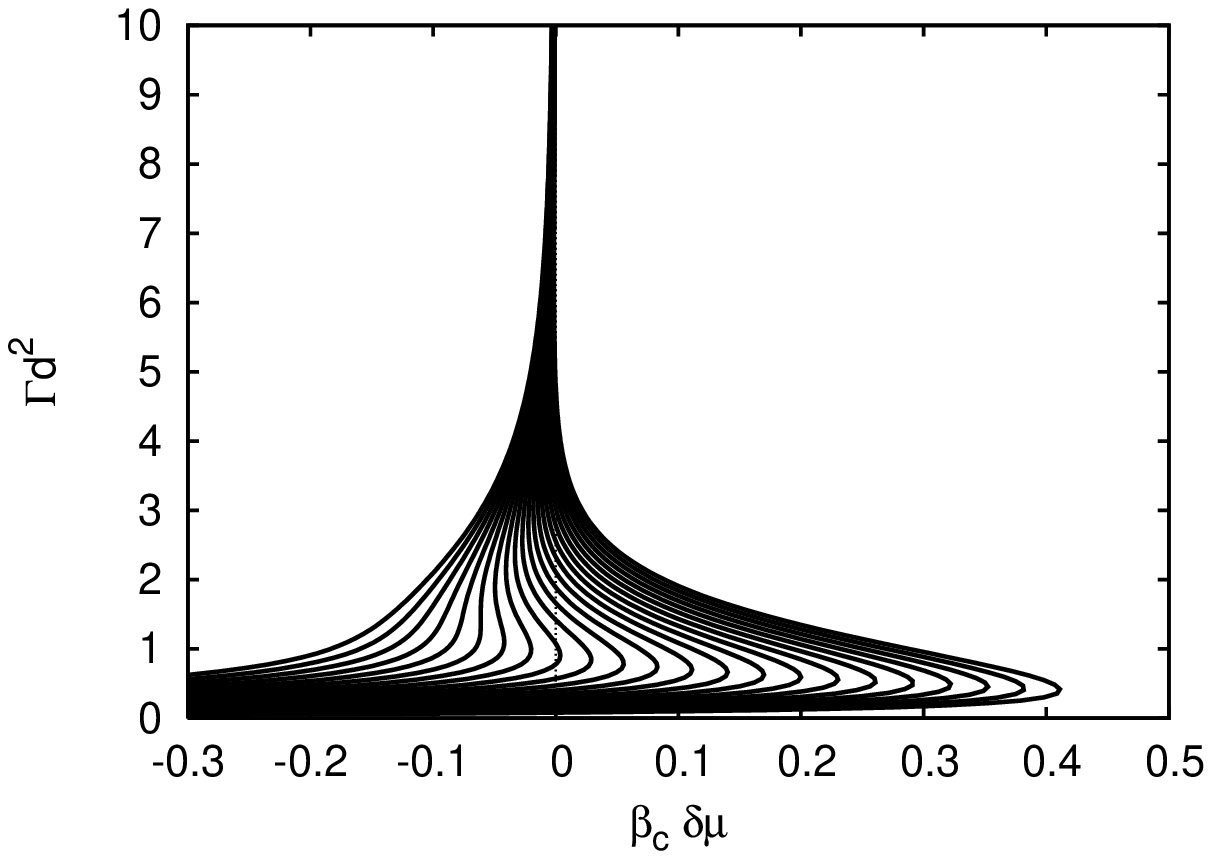}
\caption{\label{bifalg}%
Bifurcation diagram for $\Gamma$ as a function of
$\delta\!\mu\equiv\mu-\mu_{\text{sat}}$, for the attractive wall
with potential $V_{\text{LR}}$ (equation~(\ref{vextalg})),
$z_{\text{w}}=0$ and $\sigma_{\text{w}}=1.25\,d$. The temperature is
$T=0.7\,T_{\text{c}}$ and the dividing interface is located at
$z_{\text{I}}=z_{\text{w}}$. The wall parameter
$\beta_{\text{c}}\epsilon_{\text{w}}$ varies from 1.2 (right curve)
to 3.4 (left curve) in steps of 0.1.}
\end{center}
\end{figure}

\begin{figure}
\begin{center}
\includegraphics[width=0.8\linewidth]{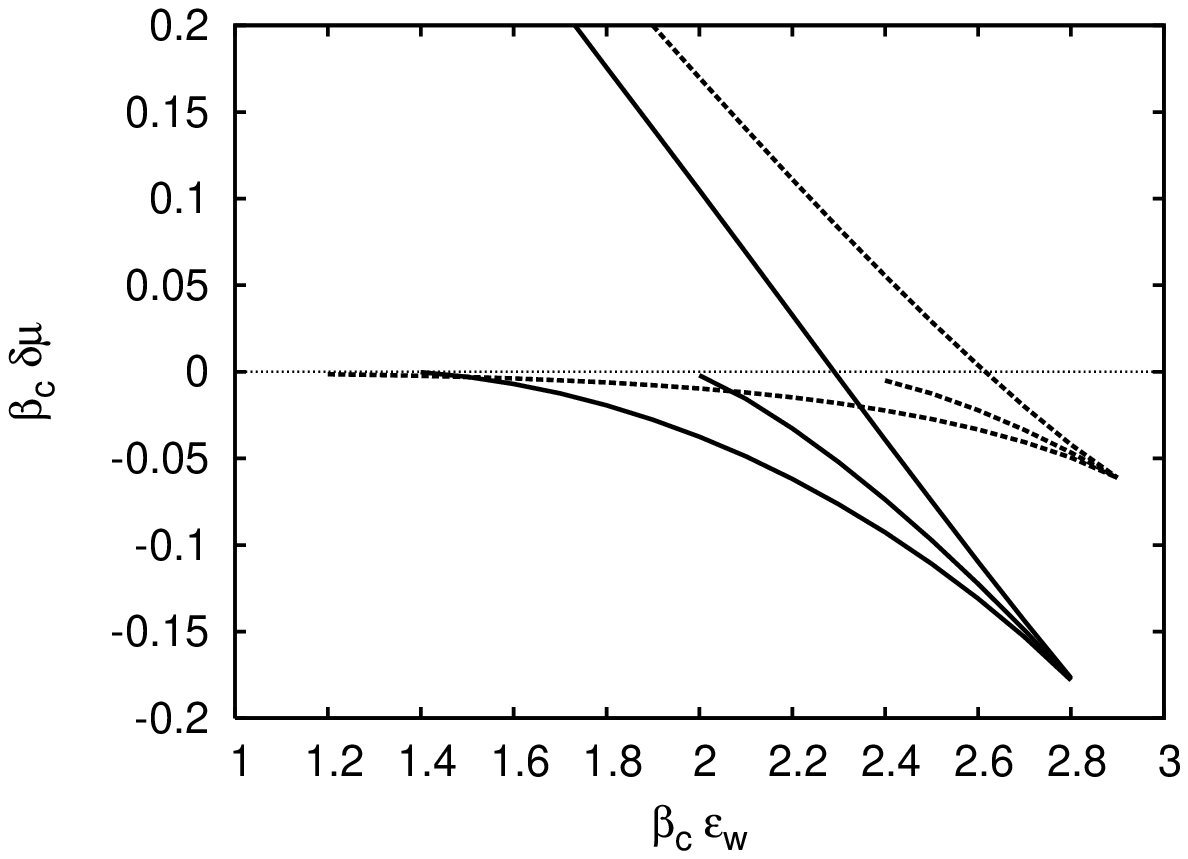}
\caption{\label{bifurc}%
Special values ($\mu_{-}$, $\mu_{\text{coex}}$ and $\mu_{+}$) of the
chemical potential as a function of $\epsilon_{\text{w}}$. The solid
lines correspond to wall potential $V_{\text{SR}}$ and the dotted
lines to wall potential $V_{\text{LR}}$. The remaining parameters
are the same as in figures~\ref{bifexp} and~\ref{bifalg}.}
\end{center}
\end{figure}

Typical isotherms for the wall potentials $V_{\text{SR}}$ and
$V_{\text{LR}}$ in equations~(\ref{vextexp}) and~(\ref{vextalg}) are
shown in figures~\ref{bifexp} and~\ref{bifalg}, respectively. The
isotherms appear qualitatively similar to those in
figure~\ref{biflj}. The effect of the potential ranges, however,
becomes apparent in the upper half of the figures as, for a given
isotherm, $\mu$ tends to $\mu_{\text{sat}}$ as $\Gamma d^2$ goes to
$+\infty$ much faster in the case of potential $V_{\text{SR}}$ than
in the case of potential $V_{\text{LR}}$. It may also be worthwile
to note the important differences in the bottom half of the figures
even though the distance from the wall is small there (i.e.\ the
potential range is less critical): the values of $\mu_{-}$,
$\mu_{\text{coex}}$ and $\mu_{+}$ in the two cases differ
significantly. This is confirmed in figure~\ref{bifurc} where their
dependance on the strength $\epsilon_{\text{w}}$ of the wall
potentials is depicted. The intersection of $\mu_{\text{coex}}$ with
the $x$-axis (i.e.\ the $\mu=\mu_{\text{sat}}$ axis) in
figure~\ref{bifurc} corresponds to the partial/complete wetting
transition with respect to $\epsilon_{\text{w}}$. For weakly
attractive walls ($\epsilon_{\text{w}}$ small), the wetting is
partial and becomes complete for very attractive walls
($\epsilon_{\text{w}}$ large). In the present case, the transition
occurs sooner for the short-range potential.

Note that independently of the situation, complete wetting or
partial wetting, there is always a film (albeit very thin) in
contact with the wall. After all, the wall is always attractive. If
it were repulsive, i.e. a hard or dry wall, there would be no
contact between the fluid and the wall.

\subsection{Square gradient approach}

A substantial simplification to equation~(\ref{fngeneral}) can be
achieved if we assume that $n(\boldr)$ varies smoothly in
the range of $\phi_{\text{p}}$.
Indeed, by expanding $n(\boldr)$ in a Taylor series and
neglecting terms of $O(3)$ and higher, equation~(\ref{fngeneral})
becomes
\begin{subequations}
\begin{equation}
F[n]=F_{\text{r}}[n]+%
\int\!\dd\mathbf{r}\left(\frac{K}{2}{|\grad n(\mathbf{r})|}^2-\alpha n^2(\mathbf{r})%
\right)
\end{equation}
where
\begin{equation}
K=-\frac{2\pi}{3}\int\!\dd r\,r^4\phi_{\mathrm{p}}(r)
\text{.}
\end{equation}
\end{subequations}
Instead of the integral equation (\ref{integralequationthreedall})
where $F_{\text{r}}[n]$ is given by equation~(\ref{local}), we now
have a differential equation:
\begin{equation}
\mu_{\text{r}}(n)-K\lapl n-2\alpha n+V(\boldr)=\mu
\label{localequation}
\text{.}
\end{equation}

\begin{figure}
\begin{center}
\includegraphics[width=0.8\linewidth]{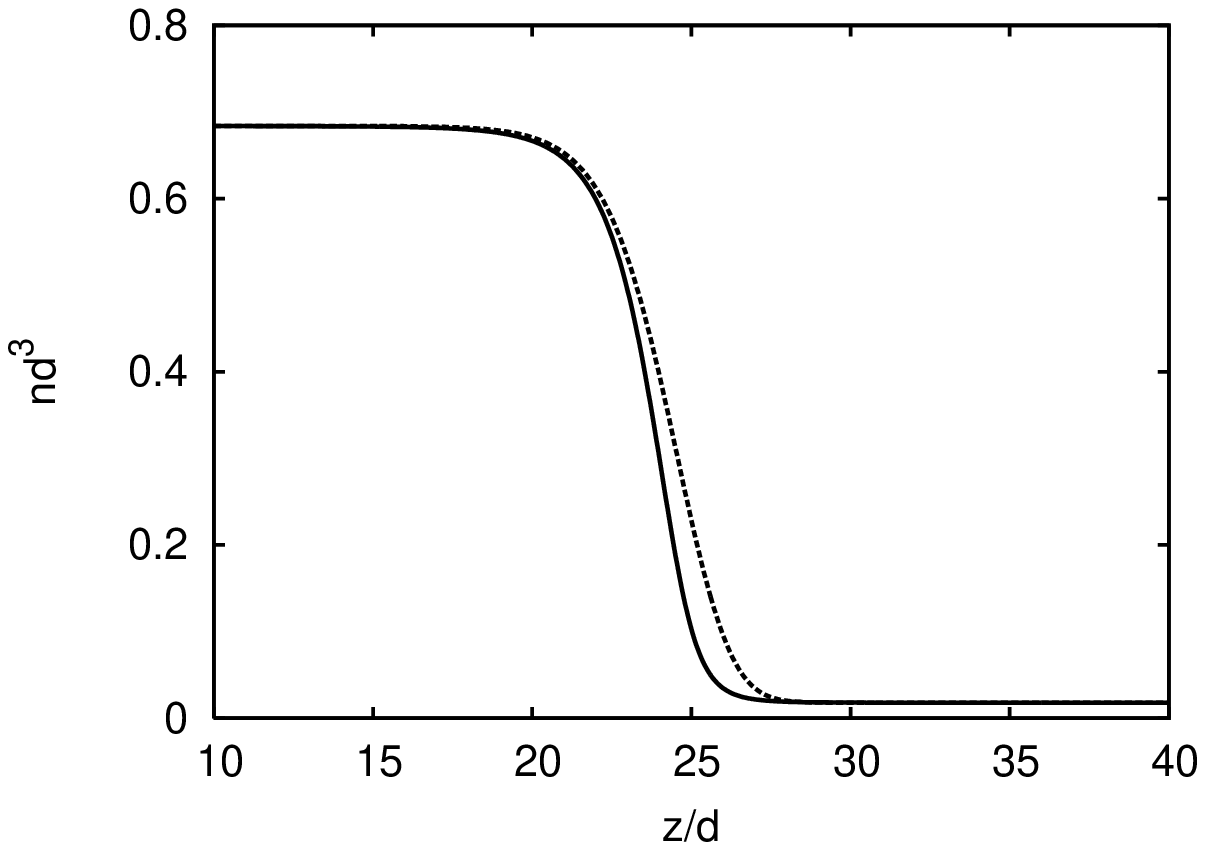}
\caption{\label{compone}%
Density profiles obtained by solving
equations~(\ref{integralequationthreedall}) (solid line) and
(\ref{localequation}) (dotted line) for $T=0.7\,T_{\text{c}}$ and
$\mu=\mu_{\text{sat}}(T)$.}
\end{center}
\end{figure}

\begin{figure}
\begin{center}
\includegraphics[width=0.8\linewidth]{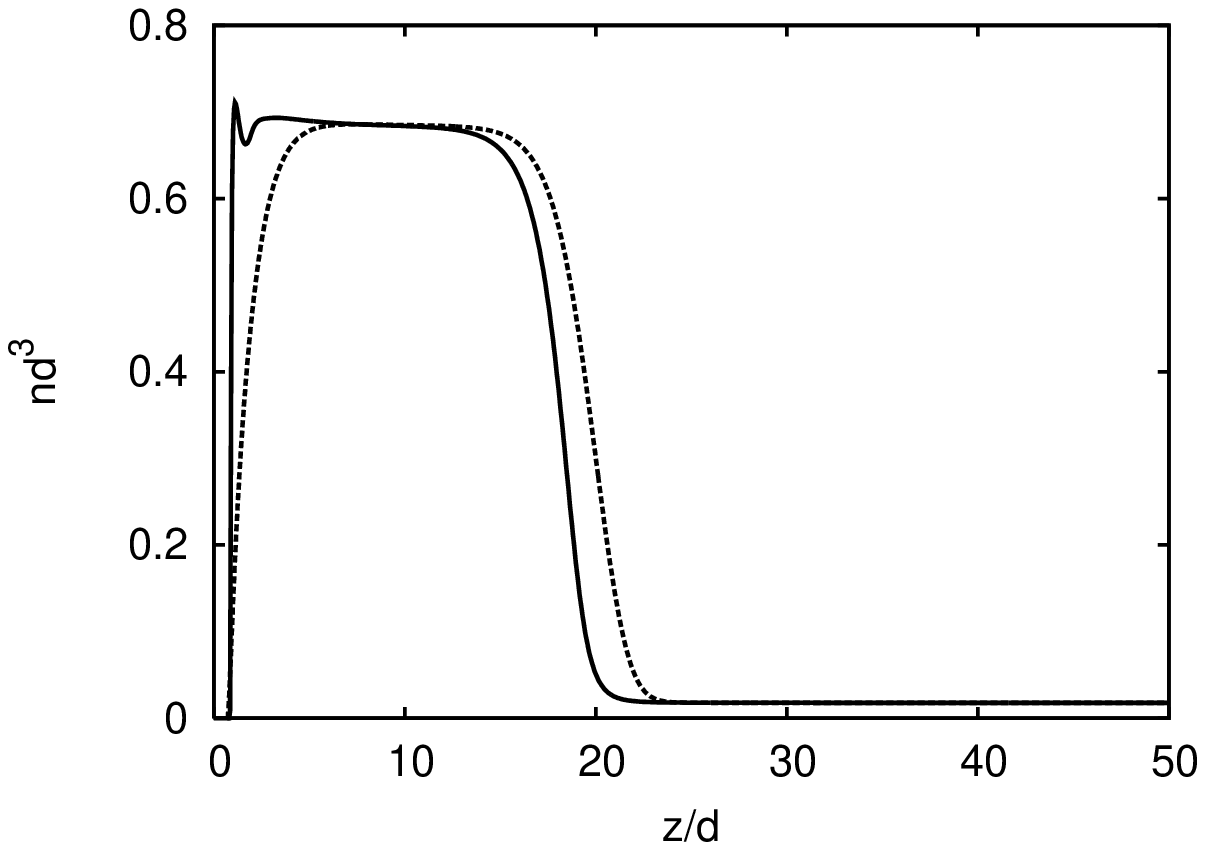}
\caption{\label{comptwo}%
Density profiles obtained by solving Eqs.~(\ref{integralequationthreedall})
(solid line) and (\ref{localequation}) (dotted line) in the presence of a wall
for $T=0.7\,T_{\text{c}}$ and $\mu-\mu_{\text{sat}}(T)=-0.001\kB T_{\text{c}}$.
The wall potential is $V_{\text{LJ}}$ with parameters
$\beta_{\text{c}}\epsilon_{\text{w}}=1.8$, $\sigma_{\text{w}}=1.25\,d$
and $z_\text{w}=0$.}
\end{center}
\end{figure}

This local approach has been widely used to compute 1D density
profiles (e.g.\ \cite{JCollInt_87_1982}). It is also the starting
point of Cahn-Hilliard-type/diffuse interface equations where the
free energy is typically a function of density and its gradient
(e.g.\ \cite{JApplMath_4_165}). However, comparison with the DFT
integral approach reveals a number of shortcomings, especially when
it comes to interfaces. Figure \ref{compone} shows the liquid/gas
interface obtained by solving
equations~(\ref{integralequationthreedall}) and
(\ref{localequation}). Although the overall shape of the density
profile is qualitatively similar for the two cases, differences
appear in the transition area, where $n(\boldr)$ varies sharply.
This is even more evident when a wall is present, as demonstrated in
figure~\ref{comptwo}. The local approach leads to smoother curves
with milder slopes and is unable to account for the small
oscillations near the wall.

\begin{figure}
\begin{center}
\includegraphics[width=0.8\linewidth]{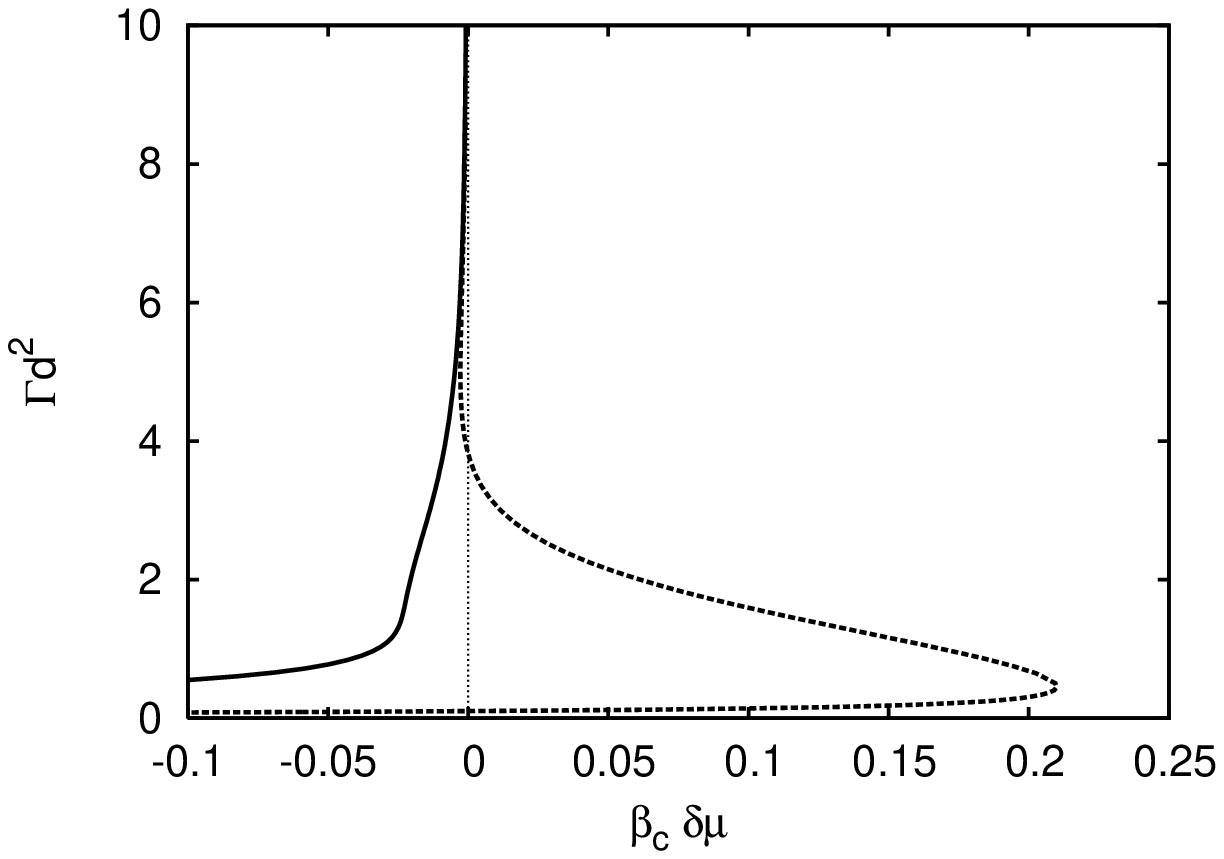}
\caption{\label{compthree}%
Isotherms obtained by solving Eqs.~(\ref{integralequationthreedall})
(solid line) and (\ref{localequation}) (dotted line) for $T=0.8\,T_{\text{c}}$.
The wall potential is $V_{\text{LJ}}$ with parameters $\beta_{\text{c}}\epsilon_{\text{w}}=1.4$,
$\sigma_{\text{w}}=1.25\,d$ and $z_\text{w}=0$.
The dividing interface is located at $z_{\text{I}}=z_{\text{w}}$.}
\end{center}
\end{figure}

Qualitative differences can appear between the two methods when
considering the isotherms. For example, in figure~\ref{compthree}
the local approach shows a multi-valued curve and hence phase
transition as opposed to a single-valued curve obtained from the
integral approach.

\section{2D problem}

\subsection{Equations}

We now consider cases for which the translational invariance along one
of the directions parallel to the substrate is broken:
\begin{subequations}
\label{integralequationtwod}
\begin{equation}
\mu_{\text{r}}[n]+%
\int\!\!\!\!\!\int_{-\infty}^{+\infty}\!\!\!\!\dd x'\dd z' n(x',z')%
\Phi_{\text{2d}}\!\left(\!\sqrt{{(x'-x)}^2+{(z'-z)}^2}\right)+V(x,z)=\mu
\end{equation}
where
\begin{equation}
\Phi_{\text{2d}}(R)=\int_{-\infty}^{+\infty}\!\!\!\dd y\,%
\phi_{\text{p}}\!\left(\sqrt{R^2+y^2}\right)
\end{equation}
with
\begin{equation}
R \equiv \sqrt{{(x'-x)}^2+{(z'-z)}^2}.
\end{equation}
\end{subequations}
The $x$-dependence of the wall potential $V$ could correspond, for
example, to a chemically heterogeneous substrate.

The integral in~(\ref{integralequationtwod}b,c) can be performed
analytically (see Appendix A). The computations now are
substantially more demanding compared to the previous ones, as
besides being 2D, they also usually require a much larger number of
iterations to achieve convergence. As in the 1D case, we utilize a
Newton scheme with an appropriately simplified Jacobian. Details are
given in Appendix B.

\subsection{Prewetting transition}
\label{prewetting}

\begin{figure}
\begin{center}
\includegraphics[width=0.8\linewidth]{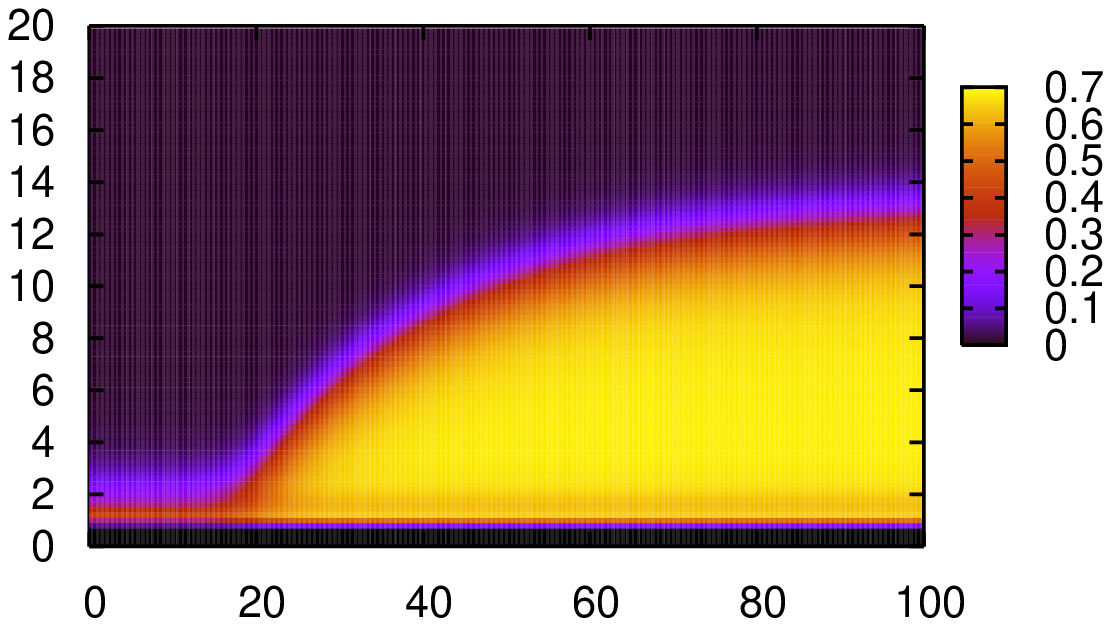}
\caption{\label{translj}%
Equilibrium density profile between two equilibrium film thicknesses
for the long-range wall potential $V_\text{LJ}$ given by
equation~(\ref{vextlj}). The temperature is $T=0.7\,T_{\text{c}}$
and the chemical potential is $\mu=\mu_{\text{coex}}$. The wall
parameters are $\beta_{\text{c}}\epsilon_{\text{w}}\approx1.5$,
$\sigma_{\text{w}}=1.25\,d$ and $z_\text{w}=0$. The shade values
provided to the right of the figure correspond to the level values
of $nd^3$.}
\end{center}
\end{figure}

As demonstrated in \S~\ref{1disotherms}, for a range of values of
temperature and chemical potential there exists a first-order phase
transition with respect to the adsorption $\Gamma$, also known as
the prewetting transition. The interface profile in the area joining
two equilibrium films thicknesses at the prewetting transition is
pictured in figure~\ref{translj} for the long-range wall potential
in equation~(\ref{vextlj}). The shape of the interface between the
two films appears to be sufficiently smooth compared to the density
profile of a liquid-gas interface as the interface between the films
is several tens of molecular diameters long, while in the latter
case it is typically only of a few molecular diameters long.

\begin{figure}
\begin{center}
\includegraphics[width=0.8\linewidth]{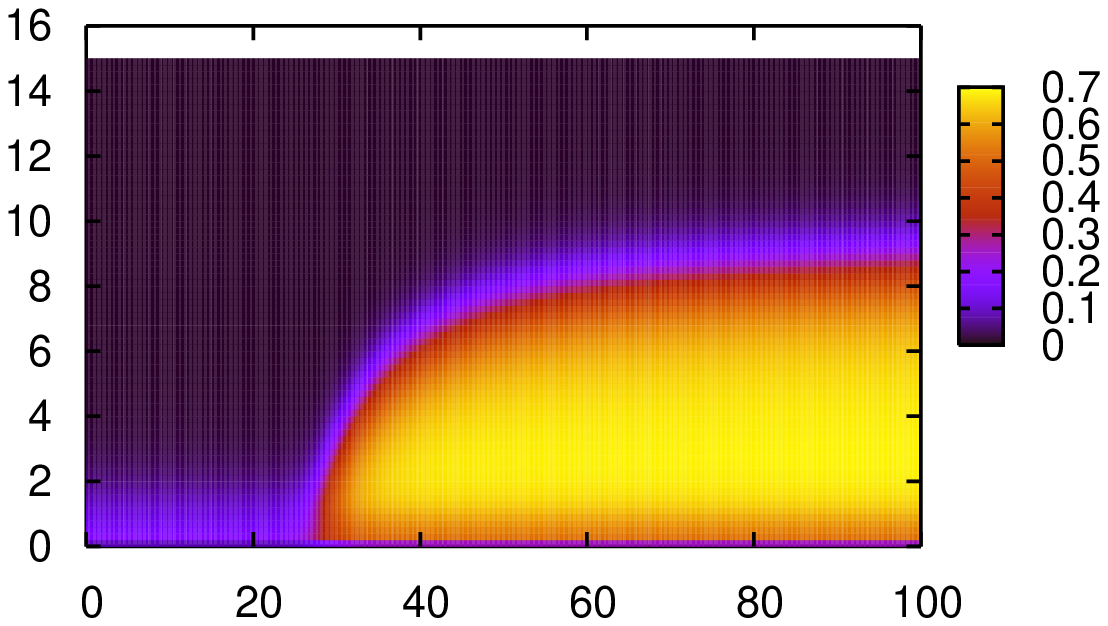}
\caption{\label{transexp}%
Equilibrium density profile between two equilibrium film thicknesses
for the short-range wall potential $V_{\text{SR}}$ given by
equation~(\ref{vextexp}). The temperature is $T=0.7\,T_{\text{c}}$
and the chemical potential is $\mu=\mu_{\text{coex}}$. The wall
parameters are $\beta_{\text{c}}\epsilon_{\text{w}}=2.0$,
$\sigma_{\text{w}}=1.25\,d$ and $z_\text{w}=0$. The shade values
provided to the right of the figure correspond to the level values
of $nd^3$.}
\end{center}
\end{figure}

The shape of the transition area depends on the difference of the
two co-existing equilibrium thicknesses and also on the range of the
wall potential. In figure~\ref{transexp}, we present the profile
obtained when the wall potential is a short-range one given in
equation~(\ref{vextexp}). The transition now between the two films
thicknesses appears to be more abrupt with a steep rim as the
smaller thickness is approached giving the profile a pancake-type
shape.

\subsection{Contact angle}

We now turn to the study of the density profile when
$\mu=\mu_{\text{sat}}$ and the wall is attractive with a potential
given by equation~(\ref{vextlj}). Recall from our discussion in
\S~\ref{1disotherms} that up to $\mu_{\text{coex}} \rightarrow
\mu^{-}_{\text{sat}}$ we can have a Maxwell construction with a thin
film co-existing with an almost infinite thick film, while at $\mu =
\mu_{\text{sat}}$ we loose the Maxwell construction and the thin
film is the most stable state. Again, as in \S~\ref{prewetting}, two
different conditions are imposed on each side of the domain. The
wall parameter $\epsilon_{\text{w}}$ is chosen so that a stable film
of finite thickness can be sustained on the wall. According to
figure~\ref{biflj}, this is true for $V_{\text{LJ}}$ when $\beta_{\text{c}}
\epsilon_{\text{w}}$ is at least in the range 0.8--1.4. The
corresponding 1D density profile computed in \S~\ref{onedpb} will
make up the boundary condition on the left side of the domain. On
the right side, a thick film of arbitrary thickness
$h_{\text{right}}$ and shape is imposed (unlike the equilibrium film
to the left, to the right we do not have an equilibrium film). For
simplicity the latter is taken to be a step function:
\begin{equation}
n_{\text{right}}(z)={}
\begin{cases}
n_{\text{liq}}\quad\text{if $z<z_{\text{w}}+h_{\text{right}}$} \\
n_{\text{gas}}\quad\text{if $z>z_{\text{w}}+h_{\text{right}}$}. \\
\end{cases}
\end{equation}

\begin{figure}
\begin{center}
\includegraphics[width=0.8\linewidth]{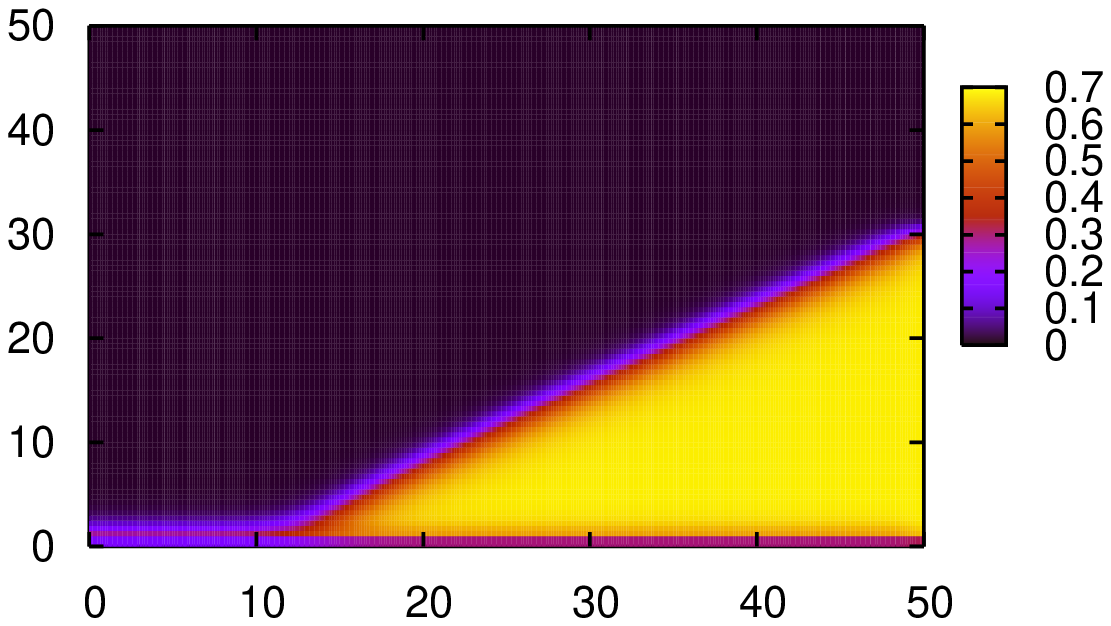}
\caption{\label{anglelj}%
Equilibrium density profile for a fluid in contact with an
attractive wall, whose potential is $V_\text{LJ}$
(equation~(\ref{vextlj})). The temperature is $T=0.7\,T_{\text{c}}$,
and the chemical potential is $\mu=\mu_{\text{sat}}$. The wall
parameters are $\beta_{\text{c}}\epsilon_{\text{w}}=1.3$,
$\sigma_\text{w}=1.25\,d$ and $z_\text{w}=0$. The contact angle is
close to $37^\circ$.}
\end{center}
\end{figure}

\begin{figure}
\begin{center}
\includegraphics[width=0.8\linewidth]{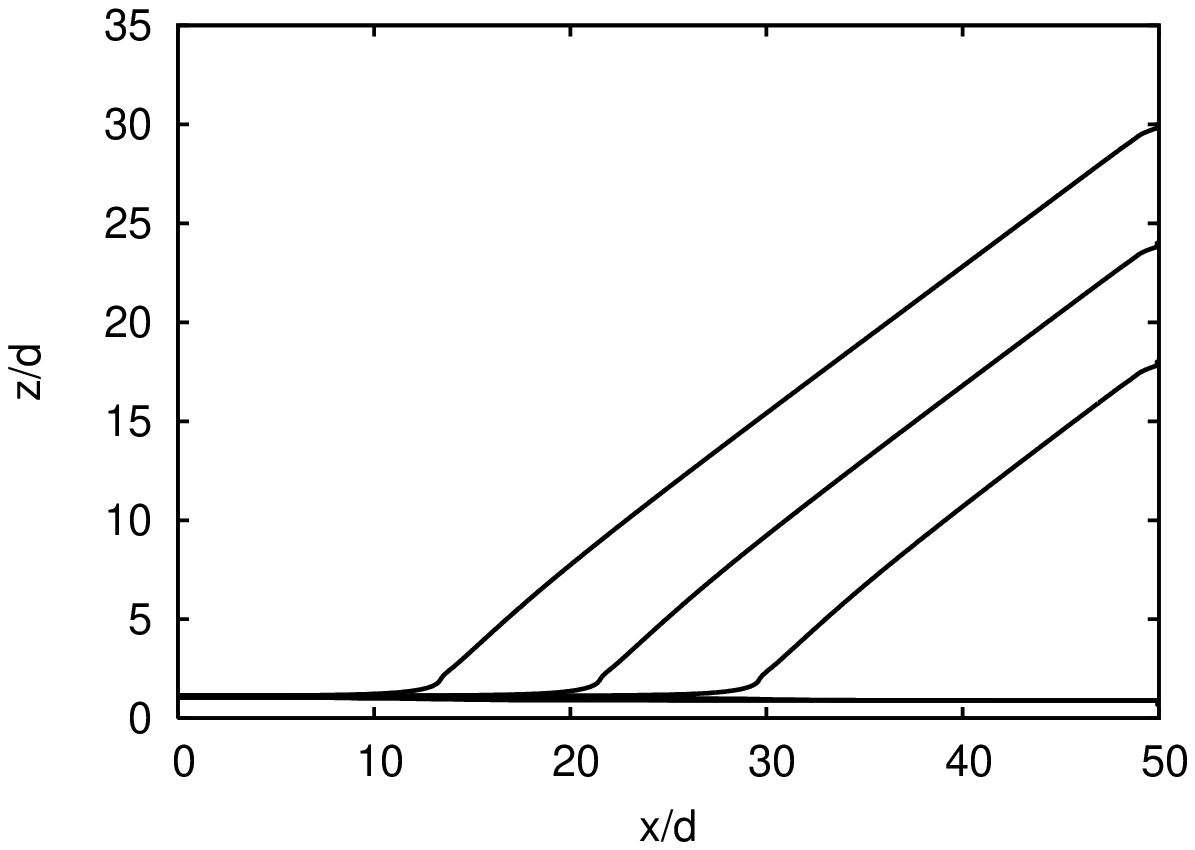}
\caption{\label{boundary}%
Density level curves (corresponding to $n=(n_{\text{liq}}+n_{\text{gas}})/2$)
for three different
values of film thickness $h_{\text{right}}$ imposed on the right
side of the domain: $h_{\text{right}}/d=18$ (right curve), $24$ and $30$
(left curve). In all cases, the same film
is imposed on the left side of the domain.
All parameters are identical to those of
figure~\ref{anglelj}.}
\end{center}
\end{figure}

An equilibrium density profile for an attractive wall obtained under
these conditions is depicted in figure~\ref{anglelj}. The main
feature is the presence of a well formed angle between the two
phases. This is more evident in the current profiles than in the
profiles obtained in \S~\ref{prewetting}, quite likely because now
the film profile on the right side of the domain is no longer an
equilibrium one. Note also the liquid film present in front of the
contact line area as the wall is attractive. Moreover, it is
important to establish that the profile is not affected by the film
thickness to the right. Indeed, as demonstrated in
figure~\ref{boundary}, the thickness of the film in the right side
of the domain has little influence on the density profile since, for
all values of $h_{\text{right}}$, the level curves are parallel to
each other sufficiently far from the contact line.

\begin{figure}
\begin{center}
\includegraphics[width=0.8\linewidth]{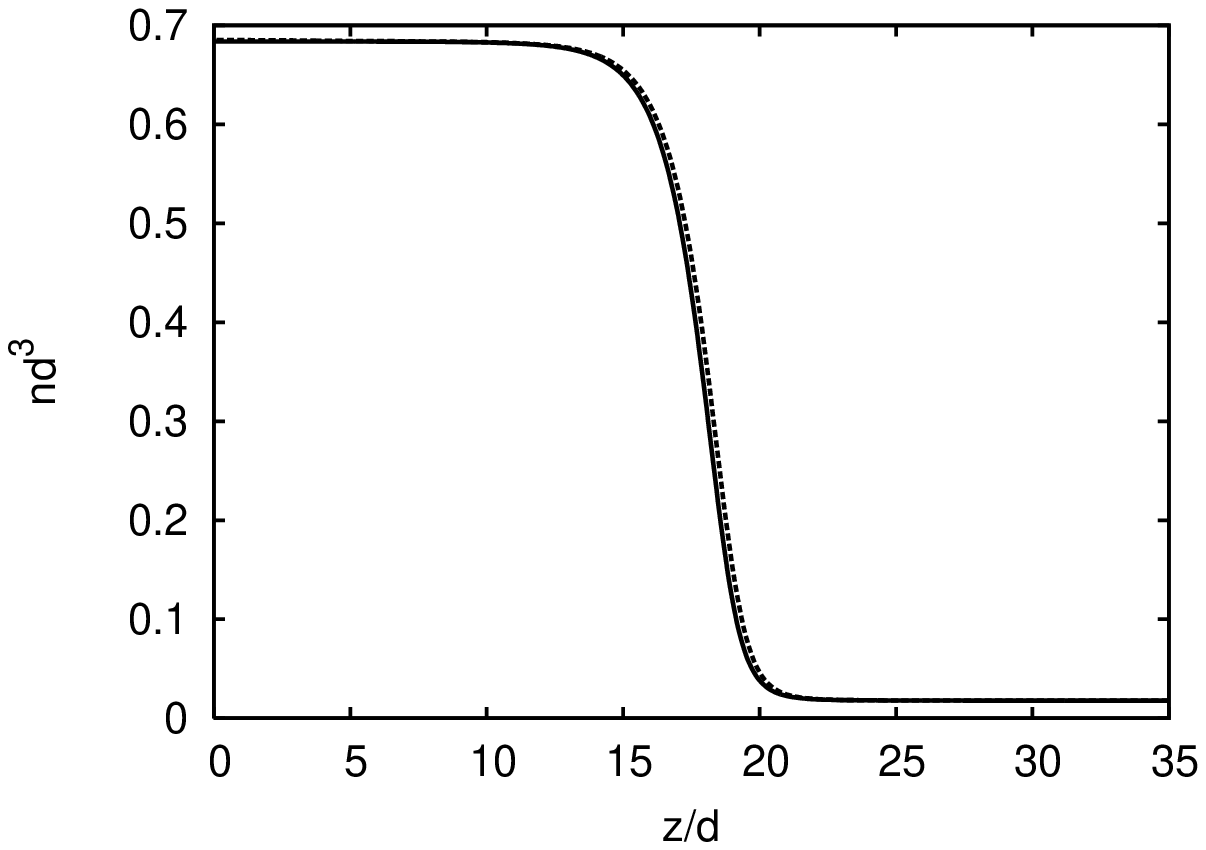}
\caption{\label{compint}%
Liquid-gas interface (solid line) from Fig.~\ref{liqgas} and cut in
the density profile pictured in Fig.~\ref{anglelj} normal to the
liquid-gas interface (dotted line). The two profiles are very close
to each other.}
\end{center}
\end{figure}

A cut in the density profile displayed in figure~\ref{anglelj} far
from the contact area and normal to the liquid-gas interface is
shown in figure~\ref{compint}. A comparison with the profile in
figure~\ref{liqgas} reveals that the two are very close to each
other: the interface sufficiently far from the contact line is
simply the usual liquid-gas interface, i.e. with no wall present. In
other words, away from the contact line, the interface shape is not
affected by the wall and it is like having the usual liquid-gas
interface there. However, a deviation from the liquid-gas interface
occurs in the three-phase area where a small incurvation of the
density profile towards the wall is observed. We shall return to
this point shortly.

\begin{figure}
\begin{center}
\includegraphics[width=0.8\linewidth]{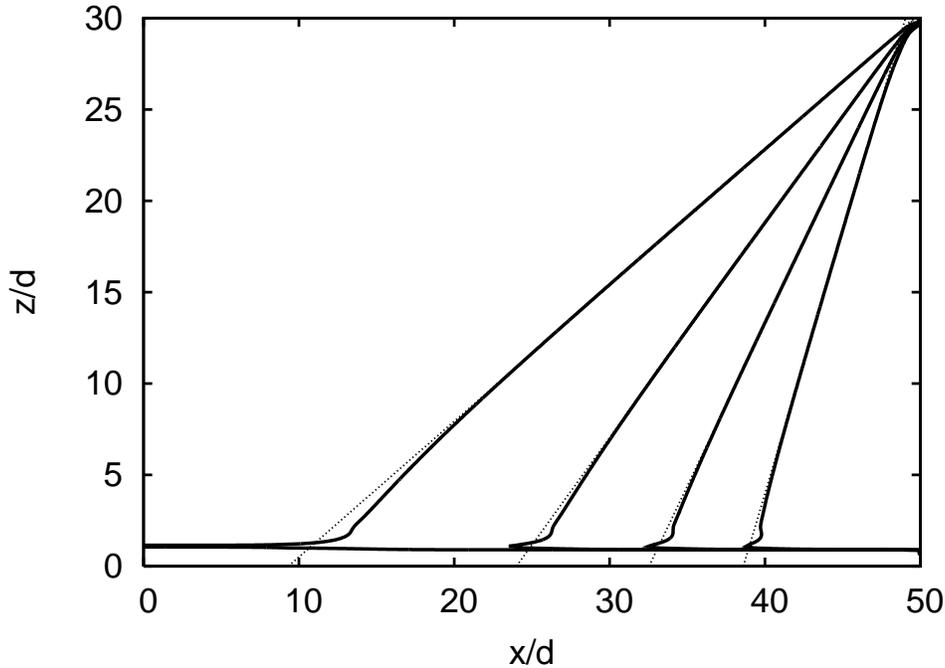}
\caption{\label{younglj}%
Density level curves corresponding to $n=(n_{\text{liq}}+n_{\text{gas}})/2$ for several
values of $\epsilon_{\text{w}}$. From right to left:
$\beta_{\text{c}}\epsilon_{\text{w}}=1.0$,
$\beta_{\text{c}}\epsilon_{\text{w}}=1.1$,
$\beta_{\text{c}}\epsilon_{\text{w}}=1.2$ and
$\beta_{\text{c}}\epsilon_{\text{w}}=1.3$. The contact
angle in each case is measured by following the dotted lines.
The wall potential is
$V_\text{LJ}$, the temperature $T=0.7\,T_{\text{c}}$
and the chemical potential $\mu=\mu_{\text{sat}}$. The wall
parameters are $\sigma_\text{w}=1.25\,d$ and $z_\text{w}=0$.
}
\end{center}
\end{figure}

The surface tension of the interface is related to the
grand potential $\Omega$ via
\begin{equation}
\sigma=\frac{\Omega-\Omega_{\text{bulk}}}{\mathcal{A}}
\end{equation}
where $\Omega_{\text{bulk}}$ is the sum of the grand potentials of
the two bulk phases and $\mathcal{A}$ is the area of the interface
under consideration. Surface tensions for the planar wall/gas,
wall/liquid and liquid/gas interfaces can be computed from the
density profile constructed in figure~\ref{anglelj}. Simple
analytical expressions can also be derived in the sharp interface
limit for an infinite system and are given in Appendix A. From the
surface tension values, one can calculate the contact angle $\theta$
from the Young equation,
\begin{equation}
\sigma_{\text{wg}}=\sigma_{\text{wl}}+\sigma_{\text{lg}}\cos\theta
\text{,}
\end{equation}
and contrast it with the one obtained by a direct geometric
measurement in figure~\ref{younglj} using density level curves. We
note that the choice of the specific level curve is not important
since, as it has been demonstrate in figure~\ref{compint}, away from
the contact line the interface shape is very close to the one
corresponding to the liquid-gas interface without wall, and
consequently two different level curves would yield the same
geometric contact angle. Results for several values of $\epsilon_w$
are given in Table \ref{surfandang}. A very good agreement is found
between the two values of the contact angle. The main discrepancy
between the two is quite likely caused by the error in the drawing
of the dotted lines in figure~\ref{younglj}. The computations in
figure~\ref{younglj} also indicate that increasing the wall
attraction decreases the contact angle (as expected).

\begin{table}
\begin{center}
\begin{tabular}{@{}c@{\hspace{8mm}}cccc@{\hspace{8mm}}c@{}}
$\beta_{\text{c}}\epsilon_{\text{w}}$ & $\beta_{\text{c}}d^2\sigma_{\text{wg}}$%
&$\beta_{\text{c}}d^2\sigma_{\text{wl}}$ & $\beta_{\text{c}}d^2\sigma_{\text{lg}}$%
&$\theta_{\text{num}} ({}^\circ)$ & $\theta_{\text{mes}} ({}^\circ)$ \\[3mm]
$1.0$ & $-0.0821$ & $-0.253$ & $0.503$ & $70.2$ & $70.8$ \\
$1.1$ & $-0.106$ & $-0.360$ & $0.503$ & $59.7$ & $60.7$ \\
$1.2$ & $-0.136$ & $-0.470$ & $0.503$ & $48.4$ & $49.5$ \\
$1.3$ & $-0.171$ & $-0.582$ & $0.503$ & $35.2$ & $36.6$
\end{tabular}
\caption{Wall/gas ($\sigma_{\text{wg}}$), wall/liquid
($\sigma_{\text{wl}}$) and liquid/gas ($\sigma_{\text{lg}}$) surface
tensions, and computed ($\theta_{\text{num}}$) and measured
($\theta_{\text{mes}}$) contact angles corresponding to
figure~\ref{younglj}.}\label{surfandang}
\end{center}
\end{table}

\subsection{Contact line}

A close inspection of the contact line area in figure~\ref{younglj}
reveals a deviation between the geometric profile suggested by
Young's equation and the one obtained from DFT close to the contact
line where the interface seems to bend towards the wall and away
from the dotted lines corresponding to Young's equation. This
feature appears to be in agreement with recent experiments
by \cite{JPhysCondMatt_17_2005}. The experiments are actually done
for a two-layer composite wall, but when the top layer is thin, the
interface close to the contact line appears to be similar to that
shown in figure~\ref{younglj}.

\begin{figure}
\begin{center}
\includegraphics[width=0.8\linewidth]{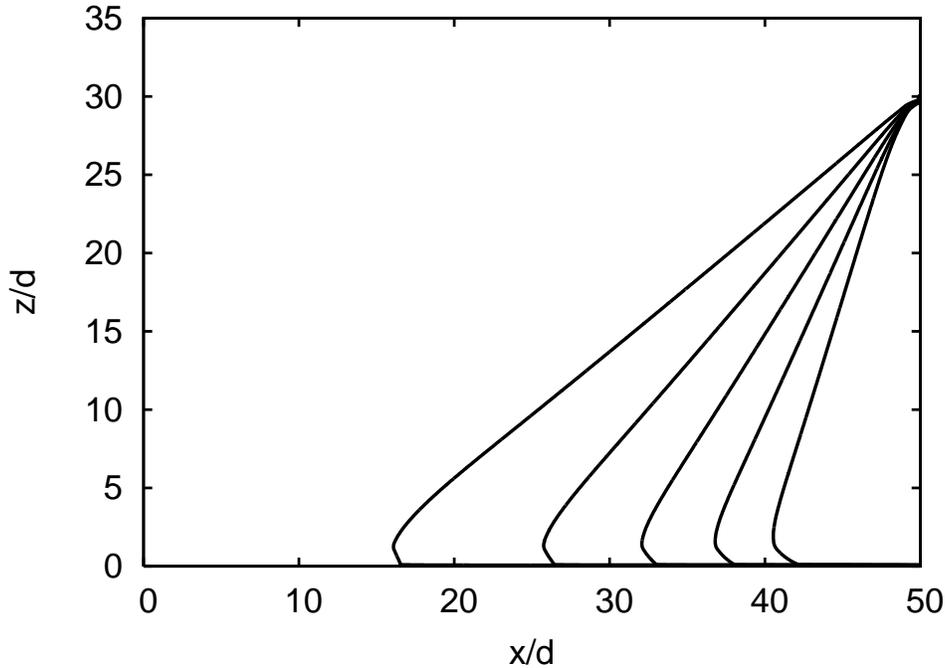}
\caption{\label{youngexp}%
Density level curves corresponding to $n=(n_{\text{liq}}+n_{\text{gas}})/2$ for wall
potential $V_{\text{SR}}$ and several values
of $\epsilon_{\text{w}}$. From right to left:
$\beta_{\text{c}}\epsilon_{\text{w}}=1.4$,
$\beta_{\text{c}}\epsilon_{\text{w}}=1.5$,
$\beta_{\text{c}}\epsilon_{\text{w}}=1.6$,
$\beta_{\text{c}}\epsilon_{\text{w}}=1.7$ and
$\beta_{\text{c}}\epsilon_{\text{w}}=1.8$.
The temperature is $T=0.7\,T_{\text{c}}$
and the chemical potential $\mu=\mu_{\text{sat}}$ while the wall
parameters are $\sigma_\text{w}=1.25\,d$ and $z_\text{w}=0$.}
\end{center}
\end{figure}

\begin{figure}
\begin{center}
\includegraphics[width=0.8\linewidth]{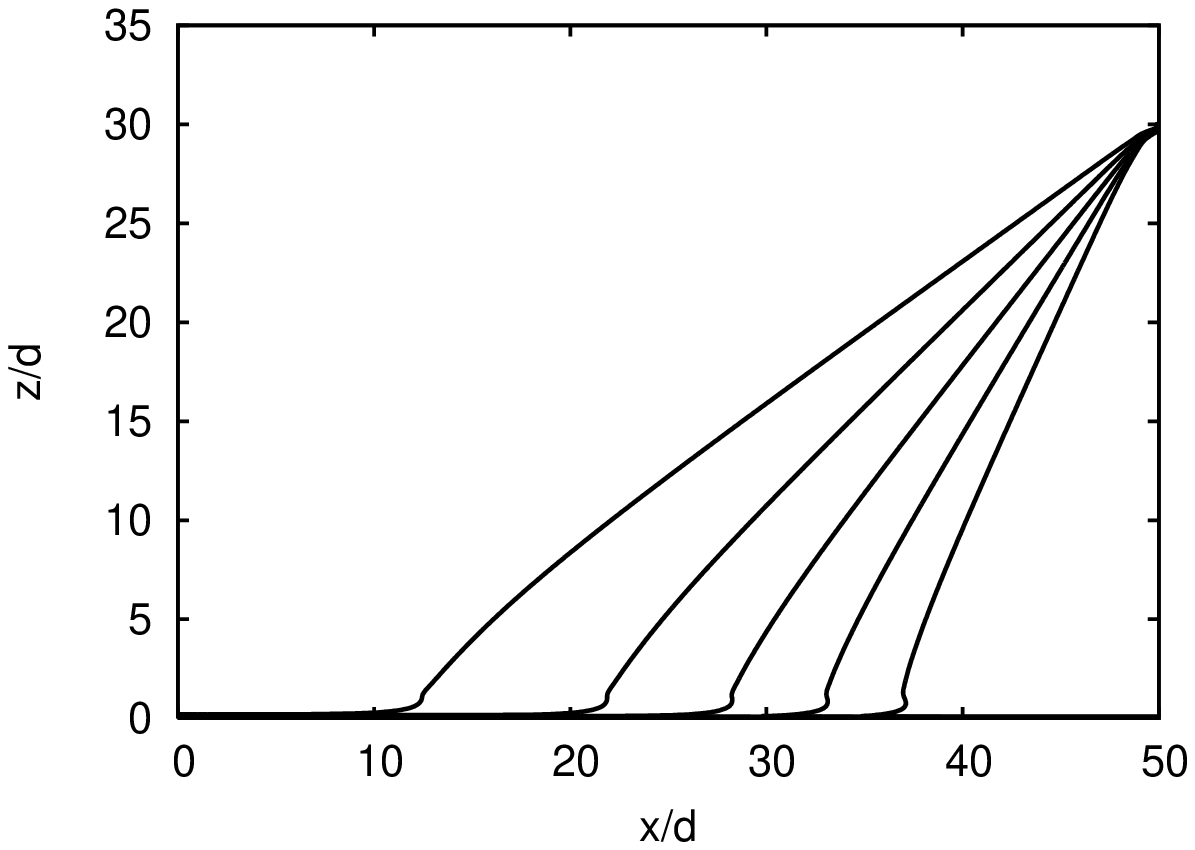}
\caption{\label{youngalg}%
Density level curves corresponding to $n=(n_{\text{liq}}+n_{\text{gas}})/2$ for wall
potential $V_{\text{LR}}$ and several values
of $\epsilon_{\text{w}}$. From right to left:
$\beta_{\text{c}}\epsilon_{\text{w}}=1.7$,
$\beta_{\text{c}}\epsilon_{\text{w}}=1.8$,
$\beta_{\text{c}}\epsilon_{\text{w}}=1.9$,
$\beta_{\text{c}}\epsilon_{\text{w}}=2.0$ and
$\beta_{\text{c}}\epsilon_{\text{w}}=2.1$.
The temperature is $T=0.7\,T_{\text{c}}$
and the chemical potential $\mu=\mu_{\text{sat}}$ while the wall
parameters are $\sigma_\text{w}=1.25\,d$ and $z_\text{w}=0$.}
\end{center}
\end{figure}

Level curves constructed with wall potentials $V_{\text{SR}}$ and
$V_{\text{LR}}$ in equations~(\ref{vextexp}) and (\ref{vextalg}),
are given in figures~\ref{youngexp} and~\ref{youngalg},
respectively. The bend is not present in the first case which
suggests that it is mainly due to the long-range nature of the
attraction of the wall potential. A parallel can be made with the
conclusions of \S~\ref{prewetting} regarding the density shape in
the transition area between the two films: as the contact angle goes
to zero, the profiles in Fig.~\ref{youngexp} seem to tend to the one
in figure~\ref{transexp} giving rise to an abrupt transition between
the two films while the ones in figure~\ref{youngalg} seem to lead
to the profile in figure~\ref{translj} for which a smoother
transition between the two films is observed. In contrast, no
dramatic difference is observed between $V_{\text{LJ}}$
(figure~\ref{younglj}) and $V_{\text{LR}}$ (figure~\ref{youngalg})
implying that the shape of the repulsive part of the wall potential
is less critical.

\begin{figure}
\begin{center}
\includegraphics[width=0.8\linewidth]{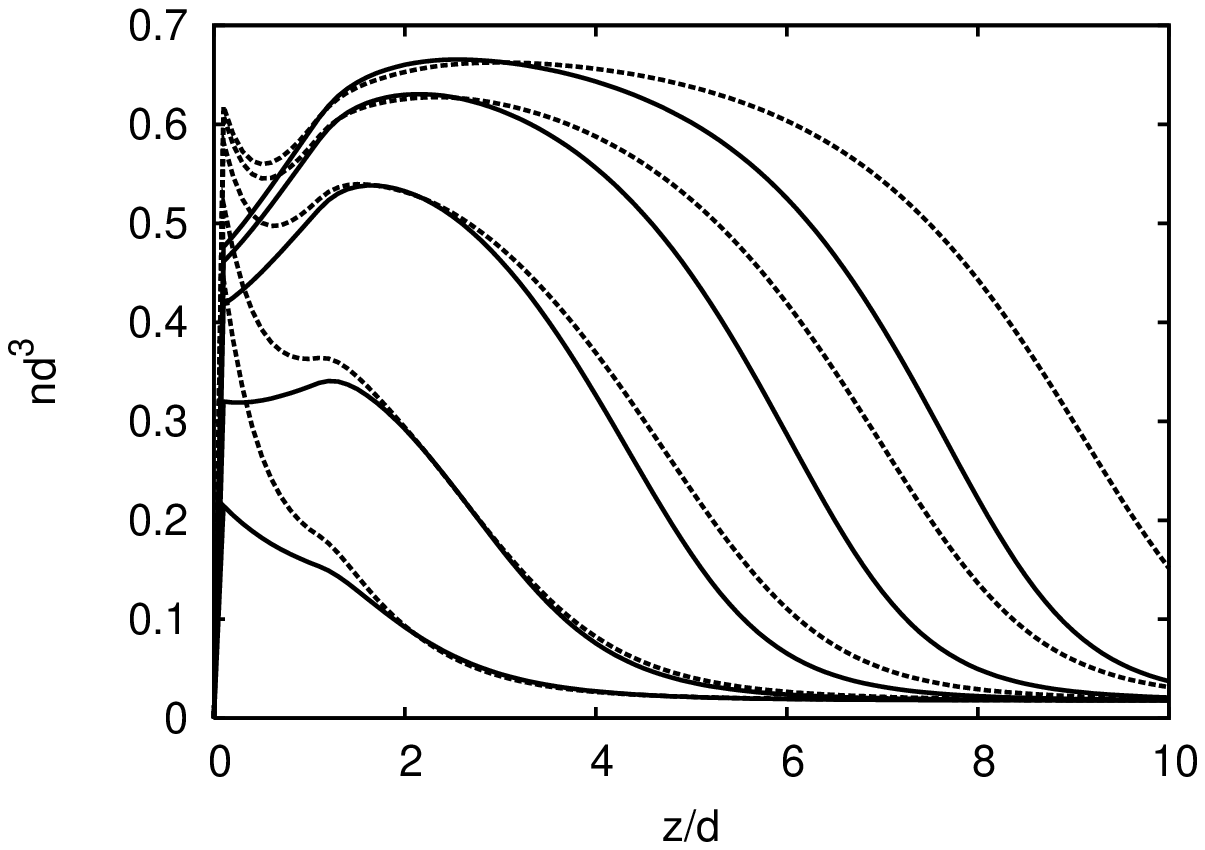}
\caption{\label{cuts}%
Density profiles around the contact line area for the cases depicted
in figure~\ref{youngexp} ($\beta_{\text{c}}\epsilon_{\text{w}}=1.8$,
solid lines) and in figure~\ref{youngalg}
($\beta_{\text{c}}\epsilon_{\text{w}}=2.0$, dotted lines).
Bottom/top curves correspond to the left/right of the contact line
region in steps of 2 for $x/d$.}
\end{center}
\end{figure}

Cuts in the contact line area normal to the wall are shown in
figure~\ref{cuts} for the two potentials $V_{\text{SR}}$ and
$V_{\text{LR}}$ in equations~(\ref{vextexp}) and~(\ref{vextalg}),
respectively. We note that the density profiles are characterized by
a small depression in the immediate vicinity of the wall ($x/d\approx
0.5$), although this feature disappears in the case of the short
range wall potential $V_{\text{SR}}$ as we move away from the
contact line towards the liquid. In contrast, it is present in the long
range case even in the 1D case as shown in figure~\ref{wallgas} for
$z\approx2d$.

\section{Summary}

We have examined the equilibrium of a fluid in contact with a solid
substrate through a DFT theory based on a perturbation approach in
which the interaction potential is split into a repulsive and an
attractive part. This leads to three distinct parts in the
expression for the free-energy functional, corresponding to
repulsion, attraction and external potential associated with the
presence of the solid boundary, respectively.

We first investigated the 1D case of a liquid film in contact with
the substrate. For values of the chemical potential less than but
close to its saturation value, a liquid film exists in contact with
the wall, even though in the absence of the wall the gas is the
preferred state. The presence of the film is due to the attractive
part of the wall potential. For a given temperature, we have
constructed bifurcation diagrams for the film thickness as a
function of the chemical potential. Such diagrams are typically
S-shaped curves with three branches. Comparison of the DFT theory
with a gradient approach in which the integral DFT equation is
approximated with a differential one reveals some significant
qualitative differences. Any gradient approach is by definition a
local one and as such it neglects the non-local features induced by
the intermolecular forces acting on the fluid system.

The bifurcation diagrams reveal that there exists a special value of
the chemical potential, which we referred to as ``coexistence"
value, in which a thin liquid film is equally stable with a thick
one (``prewetting transition"). The interface that joins the two
films is constructed with a fully 2D computation as now the
translational invariance of the system in the direction parallel to
the wall is broken. As the coexistence value tends to the saturation
one, the thickness of the thick film tends to infinity. This then
allows us to construct a liquid wedge in contact with the substrate
and with a well-defined three-phase contact line. This wedge seems
to persist for all distances from the substrate and hence it should
eventually enter the macroscale. Comparison of the contact angle
obtained from the DFT theory with a macroscopic model, such as the
Young equation based on a mechanical force balance, shows good
agreement a few molecular diameters far from the contact line.
However, unlike the Young equation which requires information on
macroscopic parameters, such as the surface tensions between the
different phases, the DFT theory relies only on well-defined
intermolecular parameters.

Of particular interest would be the extension of this approach to:
(a) Spatially heterogeneous, chemical or topographical substrates.
Such substrates have a significant effect on the wetting
characteristics of the solid-liquid pair (e.g.
\cite{JCollInt_202_1998}, \cite{PhysFluids_16_2004}, \cite{Que07},
\cite{PhysFluids_21_2009}, \cite{PhysRevLett_104_2010}); (b) The
substantially more involved dynamic case. This would form the basis
for the formulation of a dynamic contact angle theory from first
principles without the need for any phenomenological parameters. We
shall examine these problems in a future study.

\section{Acknowledgements}

We are grateful to Alexandr Malijevsky, Andreas Nold, Nikos Savva
and Petr Yatsyshin for stimulating discussions and useful comments
and suggestions. We acknowledge financial support from EPSRC
Platform Grant No. EP/E046029.

\appendix

\section{Expressions}

\subsection{Surface tensions}

For a planar interface (i.e.\ the 1D case), analytical expressions
for the surface tensions can be derived in the sharp interface limit
for an infinite system. They read:
\begin{subequations}
\begin{equation}
\sigma_{\text{lg}}=-\frac{1}{2}{(n_{\text{liq}}-n_{\text{gas}})}^2I
\end{equation}
\begin{equation}
\sigma_{\text{wg}}=-\frac{1}{2}n_{\text{gas}}^2I+n_{\text{gas}}\int_0^{+\infty}\dd zV(z)
\end{equation}
\begin{equation}
\sigma_{\text{wl}}=-\frac{1}{2}n_{\text{liq}}^2I+n_{\text{liq}}\int_0^{+\infty}\dd zV(z)
\end{equation}
where
\begin{equation}
I=\int_{-\infty}^0\dd z\int_0^{+\infty}\dd z'\Phi_{\text{1d}}(z'-z)
\text{.}
\end{equation}
\end{subequations}
Such expressions are also applicable in the contact-angle case
sufficiently far from the contact line, i.e.\ when the interface is
planar.

\vspace{0.5cm} \noindent In the following, we provide analytical
expressions used both in the 1D and 2D computations.

\subsection{Interaction potential}

The perturbative part of the interaction potential between fluid
molecules in the Barker-Henderson case is:
\begin{equation}
\phi_{\text{p}}(r)={}
\begin{cases}
\displaystyle
0\quad&\text{if $r<\sigma$} \\
\displaystyle
4\epsilon\left[{\left(\frac{\sigma}{r}\right)}^{12}-{\left(\frac{\sigma}{r}\right)}^6\right]%
\quad&\text{if $r>\sigma$}.
\end{cases}
\label{appinterpot}
\end{equation}

\subsubsection{Derivation of $\Phi$}

In the 1D case, $\Phi$, which enters the equation for the density,
is defined as:
\begin{equation}
\Phi_{\text{1d}}(\beta)=\int\!\!\!\!\!\int_{-\infty}^{+\infty}\!\!\!\dd x\,\dd y\,%
\phi_{\text{p}}\!\left(\sqrt{x^2+y^2+\beta^2}\right).
\end{equation}
Let us introduce $\bb=\beta/\sigma$. Using Eq.~(\ref{appinterpot}),
we obtain:
\begin{equation}
\label{1danalytical}
\Phi_{\text{1d}}={}
\begin{cases}
\displaystyle
-2\pi\epsilon\sigma^2\frac{3}{5}%
\quad&\text{if $|\bb|<1$} \\
\displaystyle
2\pi\epsilon\sigma^2\frac{1}{\bb^4}\left(\frac{2}{5}\frac{1}{\bb^6}-1\right)%
\quad&\text{if $|\bb|>1$}.
\end{cases}
\end{equation}

In the 2D case $\Phi$ is defined as:
\begin{equation}
\Phi_{\text{2d}}(\beta)=\int_{-\infty}^{+\infty}\!\!\!\dd y\,\phi_{\text{p}}\left(\sqrt{y^2+\beta^2}\right)
\end{equation}
For $|\bb|<1$ we obtain:
\begin{align}
\notag\Phi_{\text{2d}}=8\epsilon\sigma&\left[\frac{1}{\bb^{11}}
\left\{\frac{63}{256}\left(\frac{\pi}{2}-\arctan\left(\sqrt{\frac{1-\bb^2}{\bb^2}}\right)\right)\right.\right.\\
\notag&\qquad\left.{}-\sqrt{\frac{1-\bb^2}{\bb^2}}
\frac{\bb^2(128\bb^8+144\bb^6+168\bb^4+210\bb^2+315)}{1280}\right\}\\
&{}-\frac{1}{\bb^5}\left\{\frac{3}{8}\left(\frac{\pi}{2}-\arctan\left(\sqrt{\frac{1-\bb^2}{\bb^2}}\right)\right)
\left.{}-\sqrt{\frac{1-\bb^2}{\bb^2}}\frac{\bb^2(2\bb^2+3)}{8}\right\}\right],
\label{appphi2dsmall}
\end{align}
and for $|\bb|>1$:
\begin{equation}
\Phi_{\text{2d}}=\frac{3}{2}\epsilon\pi\sigma\frac{1}{\bb^5}%
\left[\frac{21}{32}\frac{1}{\bb^6}-1\right]. \label{appphi2dlarge}
\end{equation}

\subsection{External potential}

The Lennard-Jones expression of the wall potential can be obtained
by considering that the wall is made of particles which interact
with the fluid molecules via a pair potential of the Lennard-Jones
form:
\begin{equation}
v(r)=4\epsilon'_{\text{w}}\left({\left(\frac{\sigma_{\text{w}}}{r}\right)}^{12}%
-{\left(\frac{\sigma_{\text{w}}}{r}\right)}^6\right)
\end{equation}
where $r$ is the distance between a wall particle and a fluid
particle.

Let us consider the case of a planar wall. Our coordinate system has
the three unit vectors $(\mathbf{i},\mathbf{j},\mathbf{k})$ with
$\mathbf{k}$ an outward unit vector normal to the wall. The wall is
described as a continuous medium and the wall particles density,
denoted by $n_{\text{w}}$, is taken uniform. The the edge of the
wall, which is in direct contact with the fluid, is located at
$z'_w$ so that $n_{\text{w}}=0$ if $z>z'_{\text{w}}$. The overall
potential exerted by the wall at position $(x_0,y_0,z_0)$ in the
fluid domain ($z_0>z'_{\text{w}}$) is,
\begin{equation}
V(x_0,y_0,z_0)=\int_{z<z'_{\text{w}}}\dd\tau\,n_{\text{w}}v(r)
\end{equation}
where $n_{\text{w}}$ denotes the wall particles density and is
assumed constant. As the wall is invariant in the
$(\mathbf{i},\mathbf{j}$) directions, $V$ only depends on $z_0$.
Using cylindrical coordinates, we obtain,
\begin{equation}
V(z_0)=\int_0^{+\infty}\!\!\!\!\!R\dd R\int_0^{2\pi}\!\!\!\!\!\dd\theta%
\int_{-\infty}^{z'_{\text{w}}}\!\!\!\!\!\dd z\,%
n_{\text{w}}\,v\!\left(\sqrt{R^2+{(z_0-z)}^2}\right)
\end{equation}
which for $z_0>z'_{\text{w}}$ yields:
\begin{equation}
V(z_0)=4\pi\epsilon'_{\text{w}}n_{\text{w}}\sigma_{\text{w}}^3%
{\left(\frac{\sigma_{\text{w}}}{z_0-z'_{\text{w}}}\right)}^3%
\left[\frac{2}{15}%
{\left(\frac{\sigma_{\text{w}}}{z_0-z'_{\text{w}}}\right)}^6-1\right].
\end{equation}

\section{Numerical method}

\subsection{1D case}

The equation to solve reads:
\begin{equation}
\mu_{\text{r}}[n](z)+\int_{-\infty}^{+\infty}\!\dd z' n(z')%
\Phi_{\text{1d}}(z'-z)+V(z)=\mu
\end{equation}
for $n(z)$ in the domain $\mathcal{D}_{\text{1d}}=]z_0,z_{N_z}[$
with conditions $n=n_0$ for $z\leq z_0$ and $n=n_{N_z}$ for $z\geq
z_{N_z}$. In the presence of the wall we assume $z_0\geq
z_{\text{w}}$ (or $z_0\geq z'_{\text{w}}$). We introduce a uniform
mesh: $z_i=z_0+\Delta z\, i$, $i=0...N_z$. The integral is computed
by using a trapezoidal rule in the interval $[z_0,z_{N_z}]$ while
analytical expressions are utilized outside this interval, obtained
by using~(\ref{1danalytical}) and the fact that the density there is
constant ($= n_{\text{l,g}}$). We then obtain a system of $N_z-1$
nonlinear equations for the unknowns $n_i=n(z_i)$, $i=1...N_z-1$,
which are solved by using a Newton scheme. In order to speed up the
scheme, the off-diagonal elements of the Jacobian matrix are
neglected as they are smaller compared to the diagonal ones (this is
especially so far from the diagonal). The Jacobian matrix can then
be inverted analytically in each iteration step.

\subsection{2D case}

The equation to solve reads:
\begin{equation}
\mu_{\text{r}}[n](z)+%
\int\!\!\!\!\!\int_{-\infty}^{+\infty}\!\!\!\!\dd x'\dd z' n(x',z')%
\Phi_{\text{2d}}\!\left(\sqrt{\scriptstyle{(x'-x)}^2+{(z'-z)}^2}\right)%
+V(x,z)=\mu
\label{appint2d}
\end{equation}
for $(x,z)\in\mathcal{D}_{\text{2d}}$ where
$\mathcal{D}_{\text{2d}}=]x_0,x_{N_x}[\times]z_0,z_{N_z}[$. In the
presence of a wall, $z_0\geq z_{\text{w}}$ (or $z_0\geq
z'_{\text{w}}$). The conditions outside this domain are
$n(x,z)=n_{\text{l}}(z)$ if $x\leq x_0$, $n(x,z)=n_{\text{r}}(z)$ if
$x\geq x_{N_x}$, $n(x,z)=n_b$ if $z\leq z_0$ and $n(x,z)=n_t$ if
$z\geq z_{N_z}$ The functions $n_{\text{l}}$ and $n_{\text{r}}$ are
such that $n_{\text{l}}(z)=n_b$ and $n_{\text{r}}(z)=n_b$ if $z\leq
z_0$, and $n_{\text{l}}(z)=n_t$ and $n_{\text{r}}(z)=n_t$ if $z\geq
z_{N_z}$. This choice of conditions for $x\leq x_0$ and $x\geq
x_{N_x}$ allows us to examine, for example, the case of a fluid in
contact with two liquid films of different thicknesses, one on each
side.

A uniform mesh is used to compute $n(x,z)$: $x_j=x_0+\Delta x\, j$,
$j=0...N_x$, and $z_i=z_0+\Delta z\, i$, $i=0...N_z$.
Equation~(\ref{appint2d}) is solved for $n_{ji}=n(x_j,z_i)$,
$j=1...N_x-1$ and $i=1...N_z-1$. The integral appearing in the
equation is computed by dividing the integration domain into three
parts: $\mathcal{D}_{\text{2d}}$,
$\mathcal{D}'_{\text{2d}}-\mathcal{D}_{\text{2d}}$ and
$]-\infty,+\infty[\times]-\infty,+\infty[-\mathcal{D}'_{\text{2d}}$.
The domain $\mathcal{D}'_{\text{2d}}$ is defined as
$\mathcal{D}'_{\text{2d}}=]x'_0,x'_{N_x}[\times]z'_0,z'_{N_z}[$,
where $x'_0=x_0-\sigma$, $x'_{N_x}=x_{N_x}+\sigma$,
$z'_0=z_0-\sigma$ and $z'_{N_z}=z_{N_z}+\sigma$. Outside
$\mathcal{D}'_{\text{2d}}$, Eq.(\ref{appphi2dlarge}) is used for any
$n_{ji}\in\mathcal{D}_{\text{2d}}$ and analytical expressions can be
derived for the integral when $|x',z'|\rightarrow\infty$ where the
density is constant ($=n_{\text{l,g}}$). In
$\mathcal{D}'_{\text{2d}}-\mathcal{D}_{\text{2d}}$, the integral is
computed numerically by using either Eq.~(\ref{appphi2dsmall}) or
Eq.~(\ref{appphi2dlarge}) once for all at the start of a run. In
$\mathcal{D}_{\text{2d}}$, the integral has to be carried out
numerically at each iteration.

The discretization of Eq.~(\ref{appint2d}) leads to a set of
$(N_x-1)\times(N_z-1)$ nonlinear equations for the
$(N_x-1)\times(N_z-1)$ unknowns $n_{ji}$. It is solved by using a
Newton scheme. To speed up the computations, the Jacobian is made
sparse by neglecting contributions when $|\boldr_1-\boldr_2|$, where
$\boldr_{1,2}$ the positions of two particles, is larger than a few
molecular diameters: the non-diagonal terms of the Jacobian which
involve $\Phi_{\text{2d}}(\boldr_1-\boldr_2)$ (for the diagonal
terms $\boldr_1-\boldr_2 = {\bf 0}$) are getting small as we move
further from the diagonal since $|\boldr_1-\boldr_2|$ increases so
that $\Phi_{\text{2d}}$ decreases. By doing so, more iterations are
needed for convergence but each one is considerably faster than
inverting the full Jacobian. Finally, at each iteration step, we
have to solve a linear system involving a sparse matrix. This stage
is performed with a Gauss-Siedel method.

\bibliography{contact}

\end{document}